\documentclass[preprint,elsevier,12pt]{elsarticle}

\usepackage{amssymb}

\usepackage{lineno}

\journal{Elsevier}

\usepackage{graphicx}
\graphicspath{./}
\usepackage{tikz}
\usepackage{colortbl}
\usepackage{amsmath,mathtools,amssymb}
\usepackage{breqn}
\usepackage{float}
\usepackage[intoc]{nomencl}
\makenomenclature
\usepackage{rotating}
\usepackage{array,multirow,makecell,varwidth}
\usepackage{siunitx}
\usepackage[utf8]{inputenc}
\usepackage{booktabs, makecell, longtable}
\usepackage{tabularx,ragged2e}
\newcolumntype{C}{>{\centering\arraybackslash}X}
\usepackage{caption}
\usepackage{lscape}

\usepackage{subfigure}
\usepackage{footnote}
\usepackage{booktabs}
\makesavenoteenv{tabular}
\makesavenoteenv{table}

\usepackage{pdfpages}
\usepackage{tabularx,ragged2e,booktabs}
\newcolumntype{L}{>{\RaggedRight\arraybackslash}X}
\usepackage{pgfplots}
\pgfplotsset{compat=1.15}
\usepackage{siunitx}
\newsavebox{\measurebox}

\usepackage[export]{adjustbox}

\usepackage{setspace}   
\usepackage{hyperref}
\usepackage{paralist}
\usepackage{empheq}
\usepackage{multirow,makecell}
\usepackage{listings}
\newcolumntype{M}[1]{>{\centering\arraybackslash}m}
\newcolumntype{D}{m{20mm}}

\usepackage{ltablex}
\usepackage{longtable}
\usepackage{textcomp}
\usepackage[printonlyused]{acronym}
\usepackage{tablefootnote}
\usepackage{cleveref}
\usepackage{todonotes}
\usepackage{enumitem}

\begin{document}

\begin{frontmatter}



\title{Simulation and evaluation of sustainable climate trajectories for aviation}

\author[label1]{T. Planès\corref{cor1}}
\ead{thomas.planes@isae-supaero.fr}
\author[label1]{S. Delbecq}
\author[label1]{V. Pommier-Budinger}
\author[label1]{E. Bénard}
\cortext[cor1]{Corresponding author (+33 5 61 33 88 51)}
\address[label1]{ISAE-SUPAERO, Université de Toulouse, 10 Avenue Edouard Belin, 31400 Toulouse, France}

\begin{abstract}
In 2019, aviation was responsible for 2.6\% of the world $CO_2$ emissions as well as additional climate impacts such as contrails. Like all industrial sectors, the aviation sector must implement measures to reduce its climate impact. This paper focuses on the simulation and the evaluation of climatic scenarios for the aviation industry. For this purpose, a specific tool (CAST for "Climate and Aviation - Sustainable Trajectories") has been developed. This tool follows a methodology adapted to aviation using the concept of carbon budget and models of the main levers of action such as the level of air traffic, reduction of aircraft energy consumption or energy decarbonisation. These models are based on trend projections from historical data or assumptions from the literature. Several scenario analyses are performed in this paper using CAST and allow several conclusions to be drawn. For instance, the modelling of the scenarios based on the ATAG (Air Transport Action Group) commitments shows that aviation would consume between 2.9\% and 3.5\% of the world carbon budget to limit global warming to 2°C and between 6.5\% and 8.1\% for 1.5°C. Also, some illustrative scenarios are proposed. By allocating 2.6\% of the carbon budget to aviation, it is shown that air transport is compatible with a +2°C trajectory when the annual growth rate of air traffic varies between -1.8\% and +2.6\%, depending on the considered technological improvements. However, in the case of a +1.5°C trajectory, the growth rate would have to be reduced drastically. Finally, analyses including non-$CO_2$ effects compel to emphasize the implementation of specific strategies for mitigating contrails.
\end{abstract}

\begin{keyword}
Sustainable aviation \sep Sustainable trajectories \sep Carbon budget \sep CAST
\end{keyword}

\nomenclature{$ASK$}{Available Seat Kilometer}
\nomenclature{$ATAG$}{Air Transport Action Group}
\nomenclature{$BC$}{Black Carbon}
\nomenclature{$BECCS$}{Bio-Energy with Carbon Capture and Storage}
\nomenclature{$CAST$}{Climate and Aviation - Sustainable Trajectories}
\nomenclature{$CORSIA$}{Carbon Offsetting and Reduction Scheme for International Aviation}
\nomenclature{$DAC$}{Dual Annular Combustor}
\nomenclature{$EU-ETS$}{European Union - Emissions Trading System}
\nomenclature{$ERF$}{Effective Radiative Forcing}
\nomenclature{$GHG$}{GreenHouse Gas}
\nomenclature{$GWP$}{Global Warming Potential}
\nomenclature{$ICAO$}{International Civil Aviation Organization}
\nomenclature{$IPCC$}{Intergovernmental Panel on Climate Change}
\nomenclature{$LCA$}{Life Cycle Assessment}
\nomenclature{$RMS$}{Root Mean Square}
\nomenclature{$RPK$}{Revenue Passenger Kilometer}
\nomenclature{$SLSQP$}{Sequential Least SQuares Programming}
\nomenclature{$TCRE$}{Transient Climate Response to cumulative carbon Emissions}

\end{frontmatter}

\printnomenclature


\section{Introduction}

Human activities generate GreenHouse Gas (GHG) emissions, in particular of $CO_2$ due to the combustion of fossil fuels. These various emissions, as well as other physical phenomena such as the modification of the terrestrial albedo, induce that the radiative forcing of the Earth, defined by the difference between solar irradiance absorbed and energy radiated emitted, becomes positive. This results in an increase in the global average temperature of the Earth. The consequences of these rapid and significant temperature variations are many and varied \cite{stocker2013climate}. Melting ice, rising sea levels, water stress, declining agricultural yields, heat waves or loss of biodiversity are examples, the extent of which will depend on the level of temperature anomalies. The Intergovernmental Panel on Climate Change (IPCC) studies these different questions through numerous reports such as \cite{ipcc2007physical, ipcc2013ipcc}. Due to climate change, the states that have ratified the Paris Climate Agreements \cite{schleussner2016science} have committed to limit global warming well below +2°C above pre-industrial levels and to pursue efforts to limit the increase to 1.5°C.

In order to comply with the Paris Climate Agreements, it is therefore necessary to set up compatible trajectories, particularly in terms of GHG emissions. For example, at the global level, the IPCC defines trajectories to limit global warming to 1.5°C or 2°C using the concept of carbon budgets \cite{masson2018global}. Several tools for exploring the impact of key levers of action on the reduction of GHG emissions have been proposed to simulate global trajectories easily. For instance, the En-ROADS simulator allows generating trajectories using different economic, technical and social parameters \cite{sterman2012climate}. Similarly, the Global Calculator tool can be used to generate trajectories based on energy, land and food scenarios \cite{strapasson2020modelling}. These different prospective scenarios can also be applied to specific sectors. The transportation sector is particularly interesting because of the rebound effect and the increase in travel speeds \cite{spielmann2008environmental}. For instance, transportation-specific transition scenarios are considered as in France \cite{bigo2020transports}, Nicaragua \cite{cantarero2019decarbonizing} or China \cite{zhang2020targeting}. More specifically, these analyses can also be applied to the sector of aviation.

\begin{figure}[hbt!]
    \centering
    \includegraphics[width=1\textwidth]{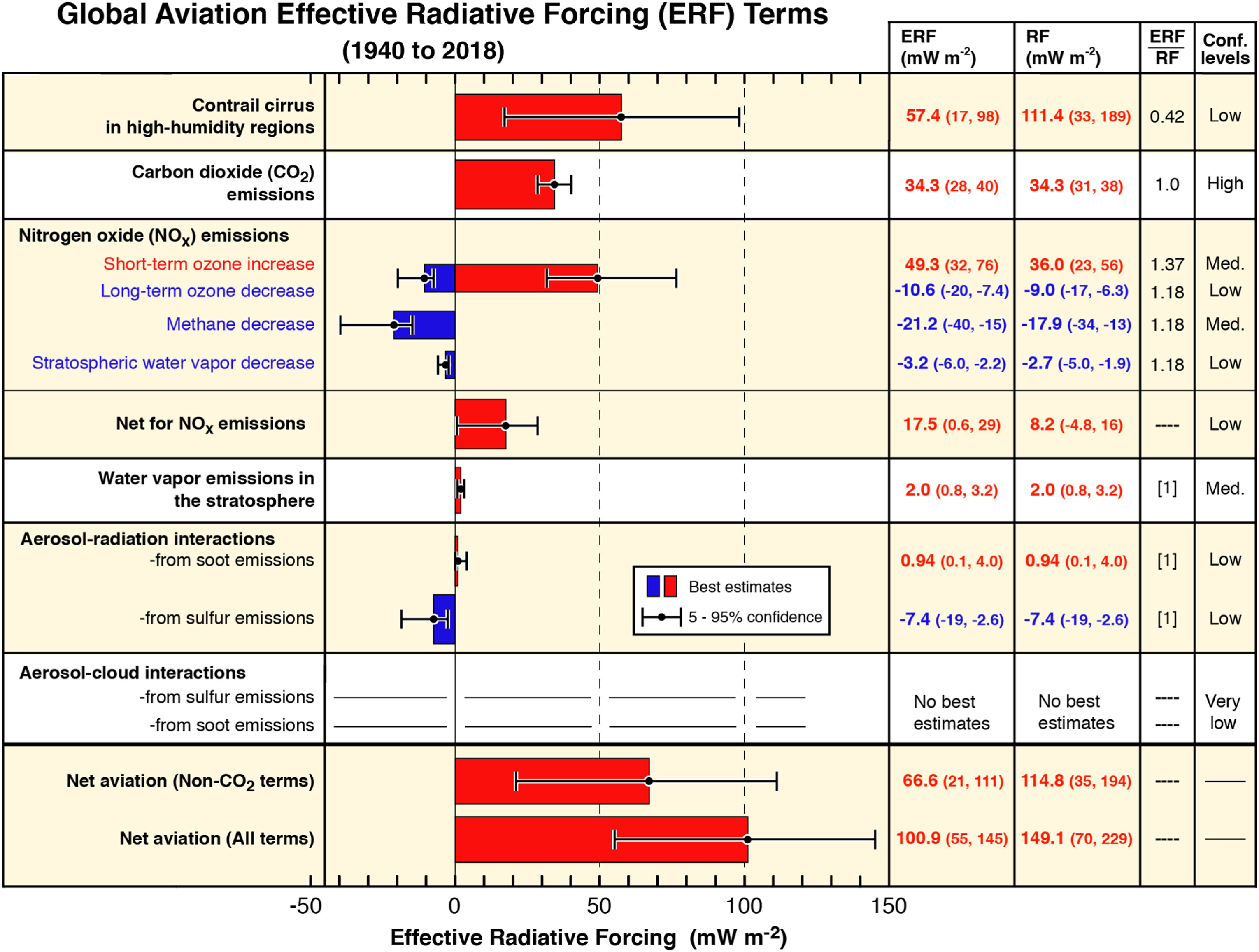}
    \caption{Different components of aviation effective radiative forcing in 2019 \cite{lee2020contribution}}
    \label{fig:aviation_RF}
\end{figure}

Aviation has a significant impact on climate change through different emissions and physical phenomena \cite{lee2009aviation}, like $CO_2$ emissions, condensation trails (contrails) or $NO_x$ emissions. It can be assessed using the notion of effective radiative forcing (ERF) \cite{ramaswamy2019radiative}. This indicator can be estimated for $CO_2$ emissions but also non-$CO_2$ effects as shown in Figure \ref{fig:aviation_RF}. Overall, aviation has generated a positive ERF of $100.9~mW/m^2$ in 2018 since 1940 and thus global warming \cite{lee2020contribution}. The non-$CO_2$ effects are dominated by contrails, which are complex phenomena that depend on local atmospheric conditions \cite{grewe2017mitigating, karcher2018formation}. From a quantitative point of view, aviation is responsible for about 2 to 3\% of world $CO_2$ emissions (2.1\% in 2019 according to \cite{atag}). In addition, by integrating the non-$CO_2$ effects like contrails, the overall climate impact of aviation reached 3.5\% of world ERF in 2011 \cite{lee2020contribution}. In addition, according to the Öko-Institut, due to the significant growth of the sector and the difficulty of easily and rapidly implementing technological solutions to reduce the GHG emissions of aircraft, the aviation sector could represent up to 22\% of global impacts on climate change by 2050 \cite{cames2015emission}. These values involve significant uncertainties, and a study is in progress to refine the results \cite{linke2020glowopt}. However, these results show that the aviation sector is responsible for significant effects on the climate and that the transition that has been initiated must be emphasized.

\begin{figure}[hbt!]
    \centering
    \includegraphics[width=0.55\textwidth]{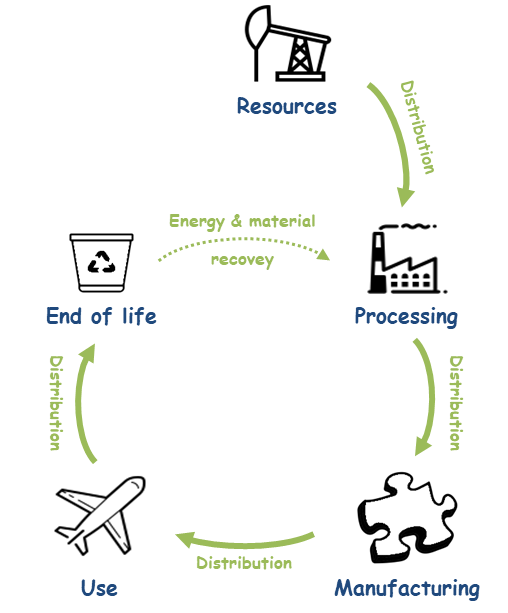}
    \caption{Life cycle for an aircraft}
    \label{fig:lifecycle_aircraft}
\end{figure}

An aircraft generates environmental impacts at different stages of its life cycle. Figure \ref{fig:lifecycle_aircraft} represents schematically the life cycle of an aircraft, including use, resource extraction or end-of-life phases. In order to better quantify the environmental impacts of aviation in the broadest sense, Life Cycle Assessment (LCA) type studies have been carried out. For example, a simplified LCA methodology for Airbus A320 aircraft has been developed \cite{johanning2014first}. A study on other aircraft has been carried out and converges towards similar results \cite{pinheiro2020sustainability}. Some studies focus more specifically on pollutant emissions close to airports \cite{kurniawan2011comparison}. All these studies show that climate impacts are one of the major environmental issues for aviation, with however some discrepancies in the evaluation of non-$CO_2$ effects. In particular, these LCAs show that the combustion and production of kerosene are the most impacting phases of the life cycle. Thus, the reduction of aircraft fuel consumption and the use of low-carbon fuels are the technological measures with the greatest impact to minimize $CO_2$ emissions from aviation. 

Numerous studies have been conducted to evaluate new technologies to reduce aircraft fuel consumption. For example, hybrid-electric architectures are being studied for aircraft with different operating ranges \cite{ribeiro2020environmental}. These architectures are envisaged for short-range aircraft. The use of new fuels is also being studied. The main solutions being considered are biofuels \cite{de2017life, zhang2020prospects} and hydrogen \cite{yilmaz2012investigation}, but both face problems of energy availability.

Knowing the climate impacts of aviation and potential improvements, work has focused on the evaluation of prospective scenarios. For instance, a 2005 study shows the need to stabilize the number of flights per inhabitant at levels slightly higher than those of the 2000s to limit the $CO_2$ atmospheric concentration to 450 ppm \cite{aakerman2005sustainable}. Moreover, the work of \cite{terrenoire2019contribution} indicates that aviation would be responsible for 5.2\% of the total anthropogenic warming under an IPCC scenario named RCP2.6, considering ICAO scenarios. Then, a study show the difficulty of decarbonising aviation \cite{sharmina2020decarbonising}. Finally, a specific economic mechanism for allocating carbon emissions is considered in \cite{qiu2017carbon} and different mechanisms such as CORSIA (Carbon Offsetting and Reduction Scheme for International Aviation) or EU-ETS (European Union - Emissions Trading System) are compared in \cite{scheelhaase2018eu}.

Although forward-looking scenarios for the aviation climate transition exist, these different studies do not address the problem in its entirety and leave open questions. First of all, non-$CO_2$ effects are often treated in an approximate way or not at all. Secondly, there are no reference models for simply constructing and analysing aviation scenarios. Thirdly, the evaluation of these scenarios with regard to the Paris Climate Agreements is not always carried out. Finally, a tool like the En-ROADS or Global Calculator tools, but dedicated to detailed simulations of the climatic trajectories specific to aviation, is missing.

The aim of the work reported here is to present a tool which can analyze scenarios for air transport with regards to environmental criteria and generate sustainable aviation trajectories in terms of climate change. The contribution of the paper is to provide models for estimating different trajectories using main aviation levers of action. The results obtained make it possible to quantify and identify general trends for aviation's climate transition and to integrate them into a single freely accessible tool.

The paper is organized as follows. In Section \ref{chap:methodology}, the overall methodology chosen for the tool is presented. Then, the models developed for estimating the impacts of aviation and assessing the sustainability of trajectories are the subject of Section \ref{chap:models}. Subsequently, in Section \ref{chap:results}, various scenarios are modelled, evaluated and criticized and a global analysis is carried out. Finally, Section \ref{chap:conclusions} offer concluding remarks and an outline of future work.

\section{Methodology}
\label{chap:methodology}

In this section, the methodology used to develop the tool CAST is outlined. First, the scope of the tool and the main data required for the implementation of the methodology followed in the tool are given. Then, the architecture of CAST is detailed as well as the main aspects of the software developments.

\subsection{Scope and data}

The scope of this work is commercial aviation which includes freight and passenger transport since freight is essentially carried out in a opportune manner (i.e. by filling the cargo compartments). In this paper, military and general aviation are not taken into account.

To develop the software, input data on global air transport are required: number of passengers, Revenue Passenger Kilometer (RPK), total aircraft distance or mean aircraft load factor. For this study, they are taken from the International Civil Aviation Organization (ICAO) \cite{icao}. The consumption of kerosene is also needed. \cite{gossling2020global} indicates that commercial aviation is responsible for the consumption of 88\% of the world's kerosene. The consumption of other fuels such as biofuels is currently marginal and is not accounted for. The world's kerosene consumption data is taken from \cite{iea}, it represents approximately 348 Mtoe in 2019. 

In order to convert this kerosene consumption into $CO_2$ emissions, European data from \cite{ademe} are used to get the emission factor estimated to $71.8~gCO_2/MJ$ if only emissions due to combustion are considered and $86.7~gCO_2/MJ$ if both kerosene production and combustion are taken into account. These values are close to the values used in American studies \cite{stratton2011impact}. To take into account the other phases of the life cycle to obtain the aviation global $CO_2$ emissions, based on mean results from \cite{pinheiro2020sustainability}, these values are increased by 2\%.

To correctly quantify the climate effects of aviation, it is necessary to also consider the non-$CO_2$ effects in addition to the $CO_2$ emissions. First, Table \ref{tab:emission_factor} gives the coefficients to obtain emissions from the consumed kerosene \cite{lee2020contribution}. To estimate the impact of these emissions in terms of ERF, coefficients are defined using data from \cite{lee2020contribution}. They are given in Table \ref{tab:erf_coef}. The impact of the contrails is estimated in relation to the total distance flown by aircraft. The impact of $CO_2$ is considered cumulative over time while the other phenomena are calculated annually. 

\begin{table}[!ht]
    \centering
    \caption{Emission factors for kerosene combustion}
    \vspace{10pt}
    \begin{tabular}{cc}
        \hline
        Emissions & Value [unit] \\
        \hline
        $CO_2$ & $3.15~[kgCO_2/kgFuel]$\\
        $H_2O$ & $1.23~[kgH_2O/kgFuel]$\\
        $NO_x$ & $15.1~[gNO_x/kgFuel]$\\
        Aerosol (BC) & $0.03~[gBC/kgFuel]$\\
        Aerosol ($SO_x$) & $1.2~[gSO_2/kgFuel]$\\
        \hline
    \label{tab:emission_factor}
    \end{tabular}
\end{table}

\begin{table}[!ht]
    \centering
    \caption{ERF coefficients for aviation climate impacts}
    \vspace{10pt}
    \begin{tabular}{cc}
        \hline
        Climate impact & Value [unit] \\
        \hline
        $CO_2$ & $0.88~[mW/m^2/GtCO_2]$\\
        $H_2O$ & $0.0052~[mW/m^2/TgH_2O]$\\
        $NO_x$ & $11.55~[mW/m^2/TgN]$\\
        Aerosol (BC) & $100.7~[mW/m^2/TgBC]$\\
        Aerosol ($SO_x$) & $-19.9~[mW/m^2/TgSO_2]$\\
        Contrails & $1.058.10^{-9}~[mW/m^2/km]$\\
        \hline
    \label{tab:erf_coef}
    \end{tabular}
\end{table}

Using all these data, direct $CO_2$ emissions from kerosene combustion for commercial aviation are computed and amount to $921~Mt$ in 2019, i.e. 2.1\% of the world $CO_2$ emissions in 2019 \cite{globalcarbon}. For comparison, ATAG has estimated these emissions at $915~Mt$ in 2019, a difference of 0.7\%. In terms of global emissions, the $CO_2$ emissions due to the entire life cycle amount to $1134~Mt$, or 2.6\% of the world $CO_2$ emissions in 2019. Including also non-$CO_2$ effects, while human activities generated $2290~mW/m^2$ until 2011 \cite{ipcc2013ipcc}, commercial aviation generated $80.6~mW/m^2$, i.e. 3.5\%.

\subsection{Architecture and development of the tool}

The objectives of CAST are to generate climate trajectories (or prospective scenarios) for aviation and to evaluate their compatibility with temperature goals such as those defined in the Paris Climate Agreements \cite{schleussner2016science}.

\begin{figure}[hbt!]
    \centering
    \includegraphics[width=1\textwidth]{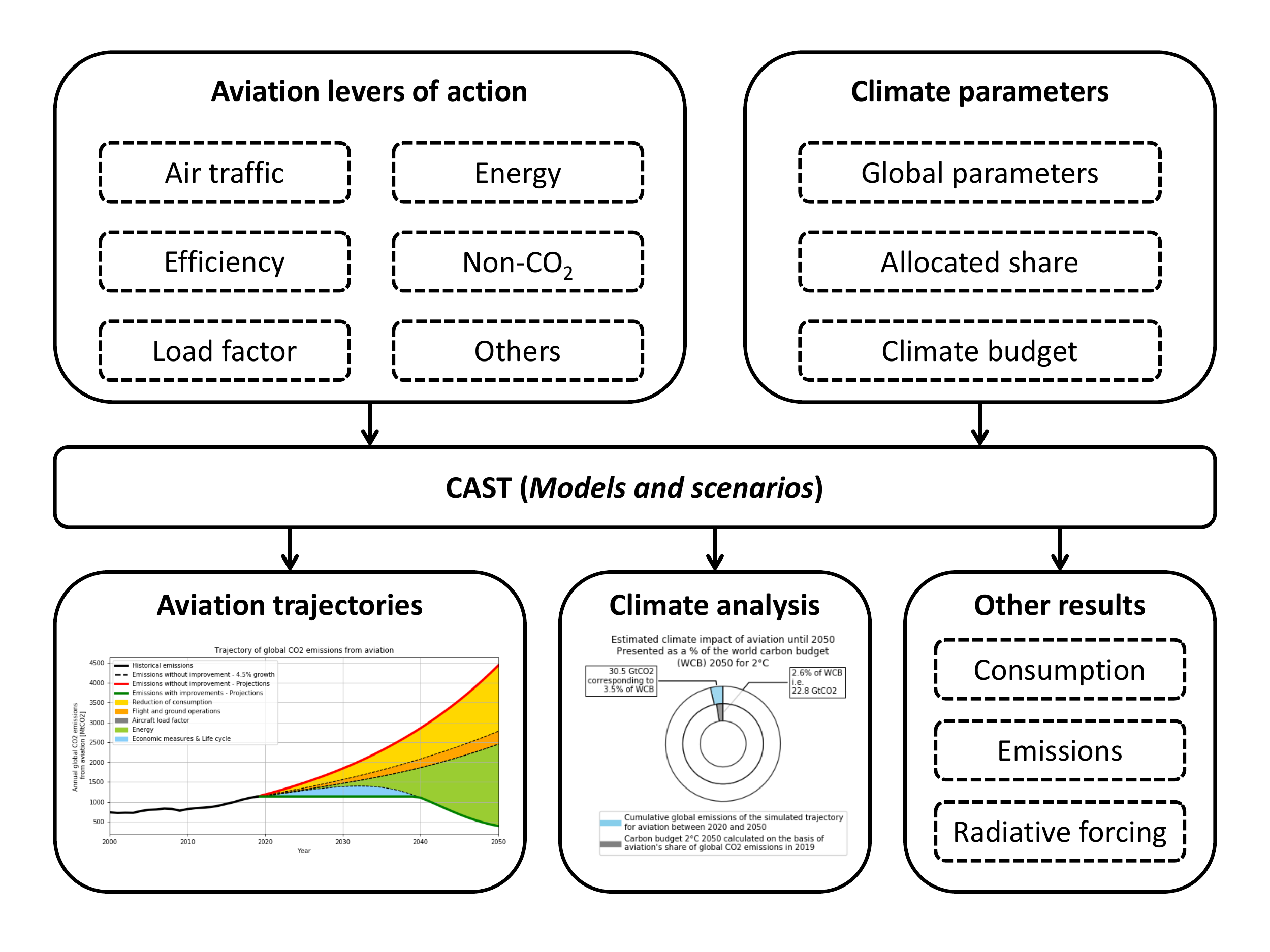}
    \caption{CAST schematic diagram}
    \label{fig:CAST_schema}
\end{figure}

Figure \ref{fig:CAST_schema} shows the schematic diagram which describes how CAST is built. CAST is based on models and scenarios, detailed in Section \ref{chap:models}, whose input data can be divided into two categories:

\begin{itemize}
    \item the main aviation levers of action such as air traffic growth or fuel consumption efficiency used in order to model the aviation sector;
    \item the climate parameters used to define climate scenario targeted for aviation.
\end{itemize}

To assess the complexity behind the CAST process, the number of inputs and outputs is given here. From its first beta-version, CAST uses 24 input variables to allow users to define their own scenarios and trajectories. In addition, it uses 69 input parameters present in the models developed to perform the analyzes proposed in CAST. These parameters are not meant to be modified by the user but rather updated when more recent literature and data are available. The CAST methodology can then compute and provide 116 outputs along with 35 different graphs.

With regard to the software development of the tool, CAST is developed using the Python programming language. The tool is freely available. Providing a free tool that scientists, organizations, authorities and companies can interact with for defining together sustainable aviation trajectories is a great motivation. The data and models are mainly manipulated and implemented using the \textit{Pandas} package \cite{mckinney-proc-scipy-2010} but also use other scientific computing package like \textit{Scipy} \cite{2020SciPy-NMeth} for solving implicit models for instance. The user interface uses \textit{ipywidgets} \cite{ipywidgets} for the widgets and \textit{ipympl} \cite{ipympl} for the graphs. The CAST software is deployed as a web application thanks to \textit{Voilà} \cite{voila}.

\section{Models}
\label{chap:models}

The purpose of this section is to present the main models used in CAST. First, the overall methodology for assessing climate trajectories is described. Subsequently, the models specific to aviation levers of action are detailed. Finally, the main climate models used are given.

\subsection{Definition of levers of action}

To simulate different scenarios of air transport, the main levers of action for aviation must be defined and interrelated. The chosen approach is based on the application of the Kaya equation to aviation. The Kaya equation (\ref{eq:kaya}) allows linking global $CO_2$ emissions to demographics (population $POP$), economics (GDP per capita $GDP/POP$) and technological parameters (energy intensity $E/GDP$ which can be related to efficiency and energy content in $CO_2$ $CO_2/E$) \cite{kaya1997environment}. The interest of this equation is that it simply shows the main levers for acting on $CO_2$ emissions \cite{friedl2003determinants}. However, some factors in the equation are interdependent and the analysis can therefore be complex \cite{schandl2016decoupling}.

\begin{equation}
\label{eq:kaya}
CO_2 = POP \times \frac{GDP}{POP} \times \frac{E}{GDP} \times \frac{CO_2}{E}
\end{equation}

Equation (\ref{eq:kaya_aviation}) is a proposal for aviation. The first factor is the Revenue Passenger Kilometer (RPK) and it represents the level of air traffic, coupling the number of passengers and the distance flown. The increase in air traffic leads to an increase in $CO_2$ emissions. The second factor $ASK/RPK$ is the ratio between the Available Seat Kilometer $ASK$ and the Revenue Passenger Kilometer $RPK$. It therefore represents the inverse of the mean aircraft load factor. For a fixed RPK, the $CO_2$ emissions decrease if the load factor increases. Next, the third factor $E/ASK$ is the ratio between the energy $E$ consumed by aviation and the Available Seat Kilometer $ASK$. It therefore represents the energy consumption per aircraft seat per kilometer and its improvement reduces $CO_2$ emissions. Finally, the last factor $CO_2/E$ is the $CO_2$ content of the energy used by the aircraft. An improvement in this factor, for example through the use of biofuels or hydrogen produced with low-carbon energy, reduces $CO_2$ emissions. These different parameters represent the main levers of action to decarbonise aviation.

\begin{equation}
\label{eq:kaya_aviation}
CO_2 = RPK \times \frac{ASK}{RPK} \times \frac{E}{ASK} \times \frac{CO_2}{E}
\end{equation}

Kaya equation for aviation being only a proposal, it can be simplified, modified or detailed. For example, additional coefficients can be added to take into account indirect emissions or non-$CO_2$ effects. Similarly, some coefficients can be refined. For instance, the factor of energy consumption per seat can be split between a factor that represents the improvements in efficiency per kilometer and another factor for improvements in flight and ground operations. Finally, it is important to note that some factors are not totally independent. For example, the fuel change may lead to an increase in energy consumption per seat or the level of air traffic may affect the mean aircraft load factor. Nevertheless, these different levers of action enable initial analyses to be carried out.

\begin{figure}[hbt!]
    \centering
    \includegraphics[width=1\textwidth]{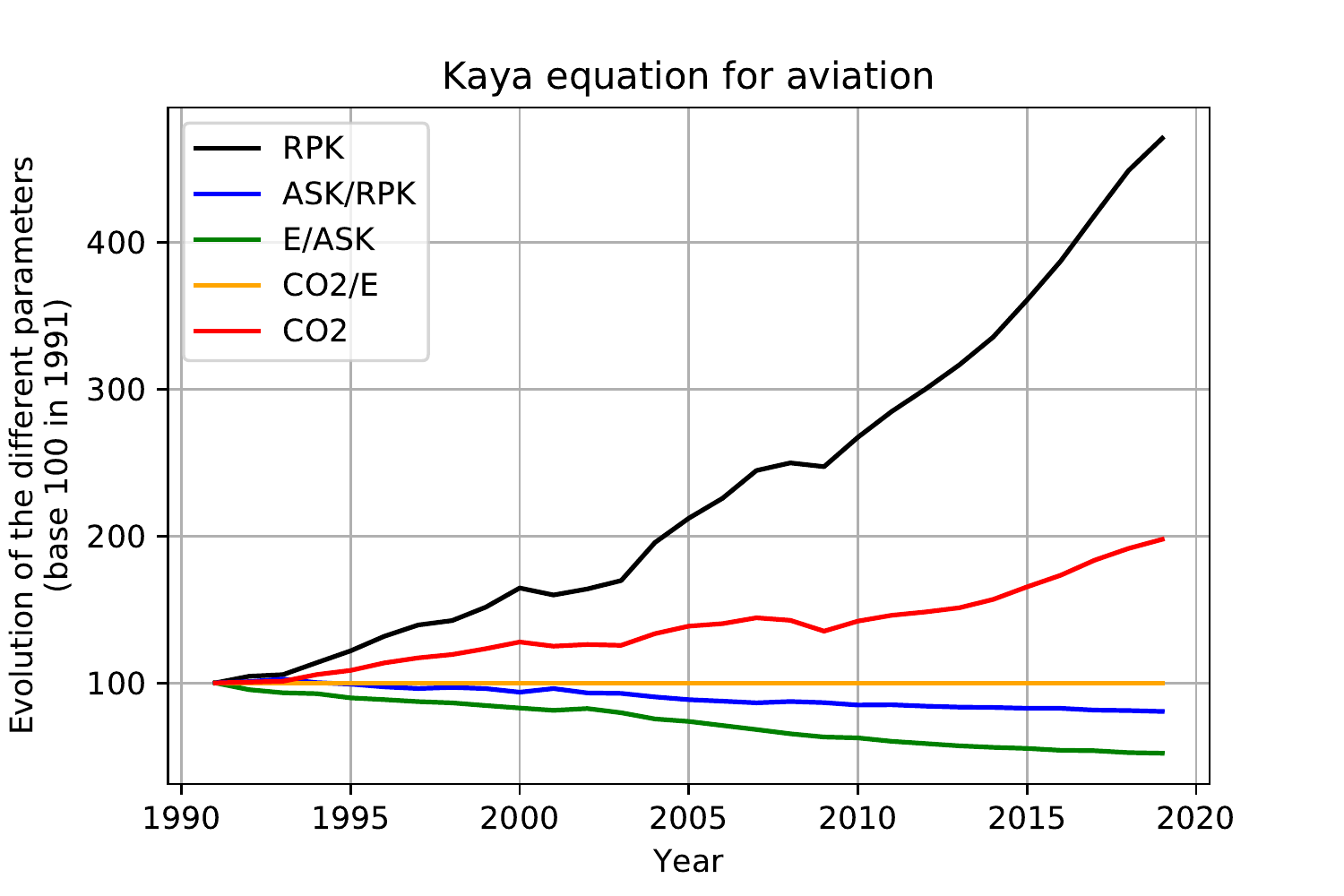}
    \caption{Evolution of Kaya equation parameters for aviation since 1991}
    \label{fig:kaya_aviation}
\end{figure}

Figure \ref{fig:kaya_aviation} represents the evolution of the different parameters from equation (\ref{eq:kaya_aviation}). Despite the improvement in the mean aircraft load factor and energy consumption per seat (divided by 2 in 30 years), aviation's $CO_2$ emissions have doubled in 30 years due to the strong increase in air traffic. It is interesting to note that due to the almost exclusive use of kerosene, the $CO_2$ energy content of aviation has remained constant. 

If the historical study of the Kaya equation makes it possible to justify the importance of the different levers of action, a projection analysis is interesting to establish transition scenarios. As a consequence, modelling the future evolution of the different parameters can allow the development of transition scenarios for aviation's $CO_2$ emissions, and more globally for the climate impact of aviation.

\subsection{Modelling the levers of action}

The objective of this section is to present the models for the various levers of action specific to aviation. The action levers chosen are those of the equation (\ref{eq:kaya_aviation}), with a distinction for operations and non-$CO_2$ effects. Two cases arise for establishing the models. Either historical data are available and deterministic historical models can be computed from these data. These models can be used to project the data into the future years to determine trend models. Or historical data is lacking and simple models are then computed on the basis of assumptions from the scientific literature.

\subsubsection{Air traffic}

\begin{figure}[hbt!]
    \centering
    \includegraphics[width=1\textwidth]{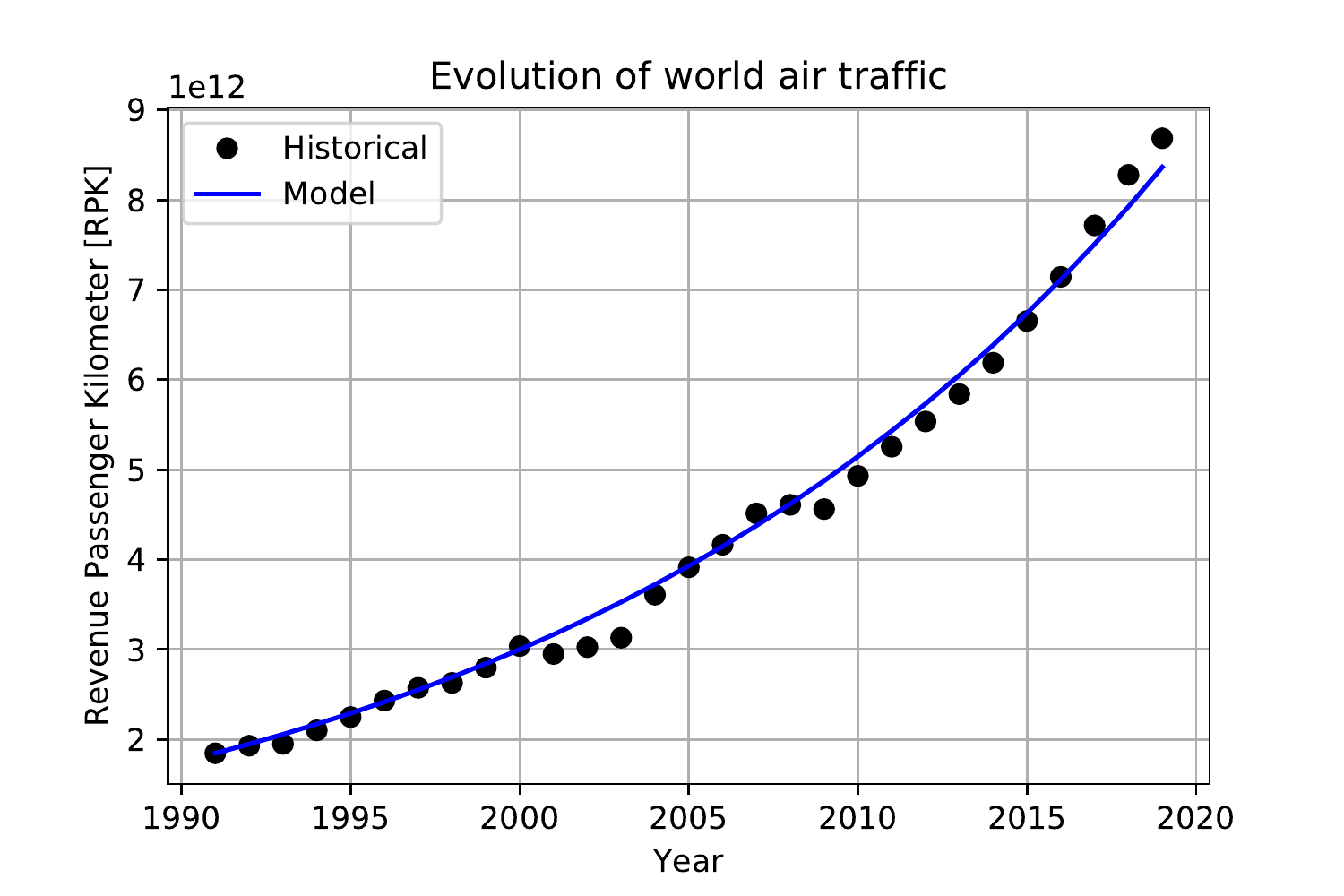}
    \caption{Model of historical world air traffic}
    \label{fig:air_traffic}
\end{figure}

The parameter corresponding to the lever of action on air traffic is $RPK$. To establish evolution scenarios, the approach consists in studying the historical evolution of this parameter. Figure \ref{fig:air_traffic} represents the historical values since 1991 \cite{icao} as well as the historical trend model. The latter was obtained using a simple exponential base function with a fixed growth rate as presented in the equation (\ref{eq:air_traffic_historical}) with $RPK_{1991}$ the initial  value in 1991, $x$ the year and $\tau$ the smoothed growth rate over the period 1991-2019.

\begin{equation}
\label{eq:air_traffic_historical}
RPK(x) = RPK_{1991} (1+\tau)^{x-1991}
\end{equation}

To determine the parameter $\tau$, an optimization was performed using the SLSQP method to minimize the Root Mean Square (RMS) error between the historical data and the model. This has the advantage of smoothing the values due to different crises (2001 September 11 attacks or financial crisis in 2008). The optimal rate obtained is then 5.5\% for the period 1991-2019, with an RMS error of 0.032. By restricting the study to the evolution of the last 10 years, this rate reaches 6.5\%, which shows an acceleration of the air traffic growth trend as depicted in Figure \ref{fig:air_traffic}.
 
Nevertheless, due to the saturation of certain markets such as Europe, manufacturers anticipate a decline in this rate in the coming years. For example, with regard to the evolution of the total distance flown by aircraft, Boeing was counting on annual growth of 4.7\% from 2017, compared with 4.4\% for Airbus \cite{fichert2020aviation}. Moreover, ICAO has announced an average forecast for RPK of 4.1\% per year between 2015 and 2045 \cite{icao_proj}. Finally, this growth rate could in the future decrease or even become negative due to the current crisis and economic, political or health measures.

To model the air traffic in the coming years, the exponential model with $\tau$ as tuning parameter is kept for its simplicity and its good representation of the evolution of this lever of action. The equation  (\ref{eq:air_traffic_model})
is used in CAST. The pre-Covid forecast growth rate is 4.5\% and the post-Covid forecast growth rate is 3.0\% \cite{atag_waypoint}.

\begin{equation}
\label{eq:air_traffic_model}
RPK(x) = RPK_{2019} (1+\tau)^{x-2019}
\end{equation}


\subsubsection{Efficiency}

The second lever of action concerns the improvement of the energy efficiency per seat, excluding the integration of flight and ground operations improvements. Contrary to air traffic trends, simple models do not adequately model historical trends. Indeed, technological limitations lead to reduced gains in recent years. For example, according to \cite{lee2010can}, energy consumption per kilometer and per passenger (including the aircraft load factor) decreased by about 1.5\% per year on average between 1975 and 2000, but less significantly afterwards. Similar results can be seen in Figure \ref{fig:kaya_aviation}.

To establish trend models of the energy efficiency per seat and scenarios, a three-step specific methodology has been developed based on historical data of energy consumption per seat from \cite{iea,icao}.

\begin{enumerate}
    \item Synthesis of a past trend model from historical data.
    \item Projection of the past trend model up to 2050 and modelling of this projection to obtain a trend model for future evolution.
    \item Definition of different scenarios using the simplified projection model
\end{enumerate}

The interest of this method is to separate the modelling of historical data from that of the projection. It allows obtaining an accurate model to represent the trend evolution and a simple model to simulate the projection and to define transition breaks. 

The difficulty is to select a type of regression model that can represent the evolution of the historical data and that allows the projection of the data in the future. Consequently, polynomial models are not considered because of their limits outside the field of study \cite{sanchez2017generation} and exponential models are preferred. 
 
\begin{figure}[hbt!]
    \centering
    \includegraphics[width=1\textwidth]{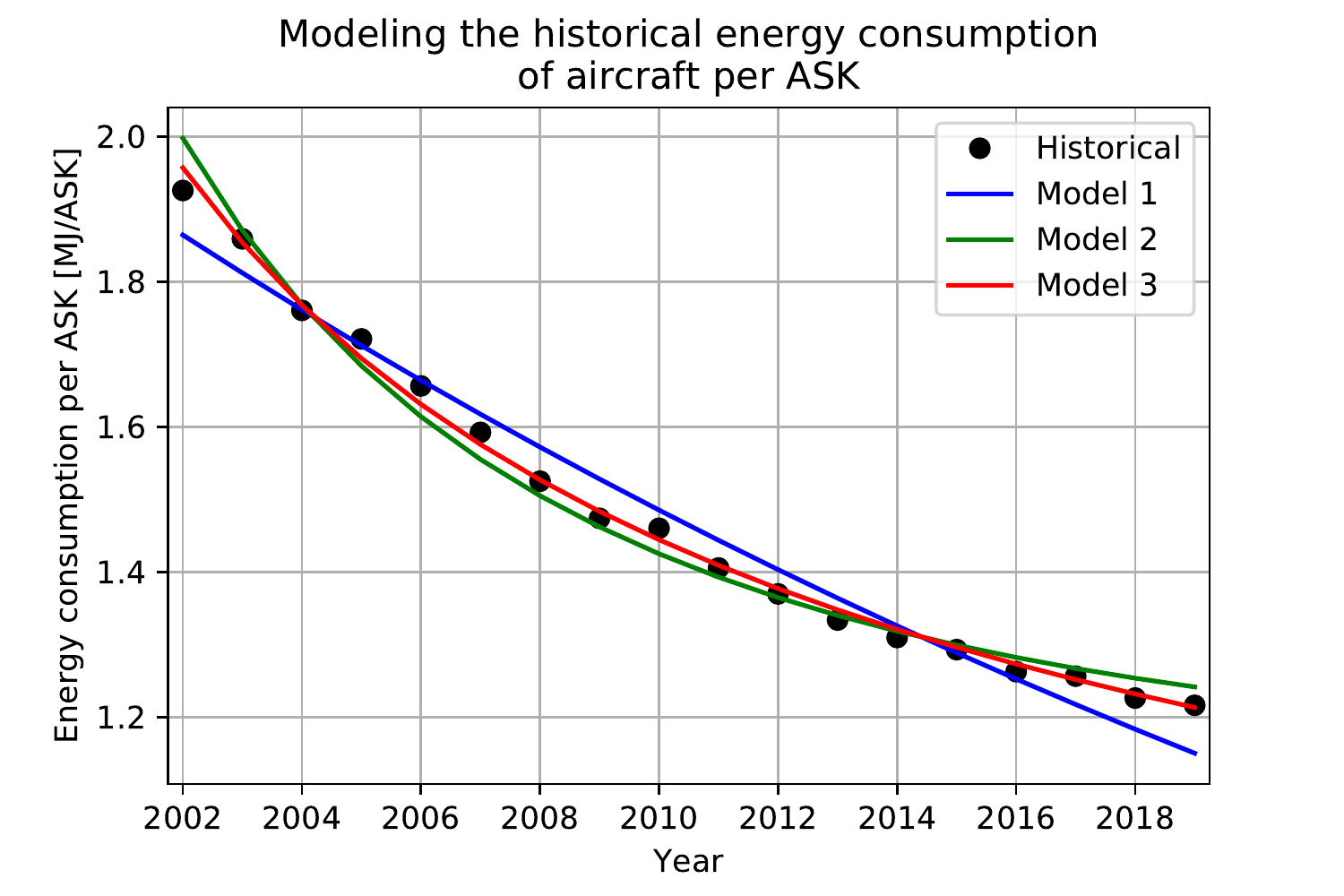}
    \caption{Models of historical aircraft energy efficiency by ASK}
    \label{fig:efficiency_hist_model}
\end{figure}

To perform the first step, three basic exponential models, more or less complex, given in the equations (\ref{eq:efficiency_f1}), (\ref{eq:efficiency_f2}) and (\ref{eq:efficiency_f3}), are here considered and compared over the period 2002-2019 due to the anomaly following 2001 September 11 attacks. For each model, an optimization using the SLSQP method is performed on the coefficients in order to minimize the RMS error between the historical data and the model. Figure \ref{fig:efficiency_hist_model} summarizes the models obtained. Model 3 provides the minimum RMS error, by a factor of 4 with respect to model 2 and by a factor of 7 with respect to model 1, which is a fixed decay rate model. Model 3 is therefore selected as the past trend model based on historical data.

\begin{equation}
\label{eq:efficiency_f1}
f_1(x) = f_0 (1-\tau)^{x-2002}
\end{equation}

\begin{equation}
\label{eq:efficiency_f2}
f_2(x) = \frac{f_f}{1-e^{-\epsilon (x-x_0)}}
\end{equation}

\begin{equation}
\label{eq:efficiency_f3}
f_3(x) = \frac{\gamma}{\beta~ln[\alpha(x-x_0)]}
\end{equation}

\noindent
with $f_0, \tau, f_f, \epsilon, x_0, \alpha, \beta, \gamma$ different coefficients. For the selected model 3: $\gamma=2.0, \beta=0.72, \alpha=0.35, x_0=1990$.

\begin{figure}[hbt!]
    \centering
    \includegraphics[width=1\textwidth]{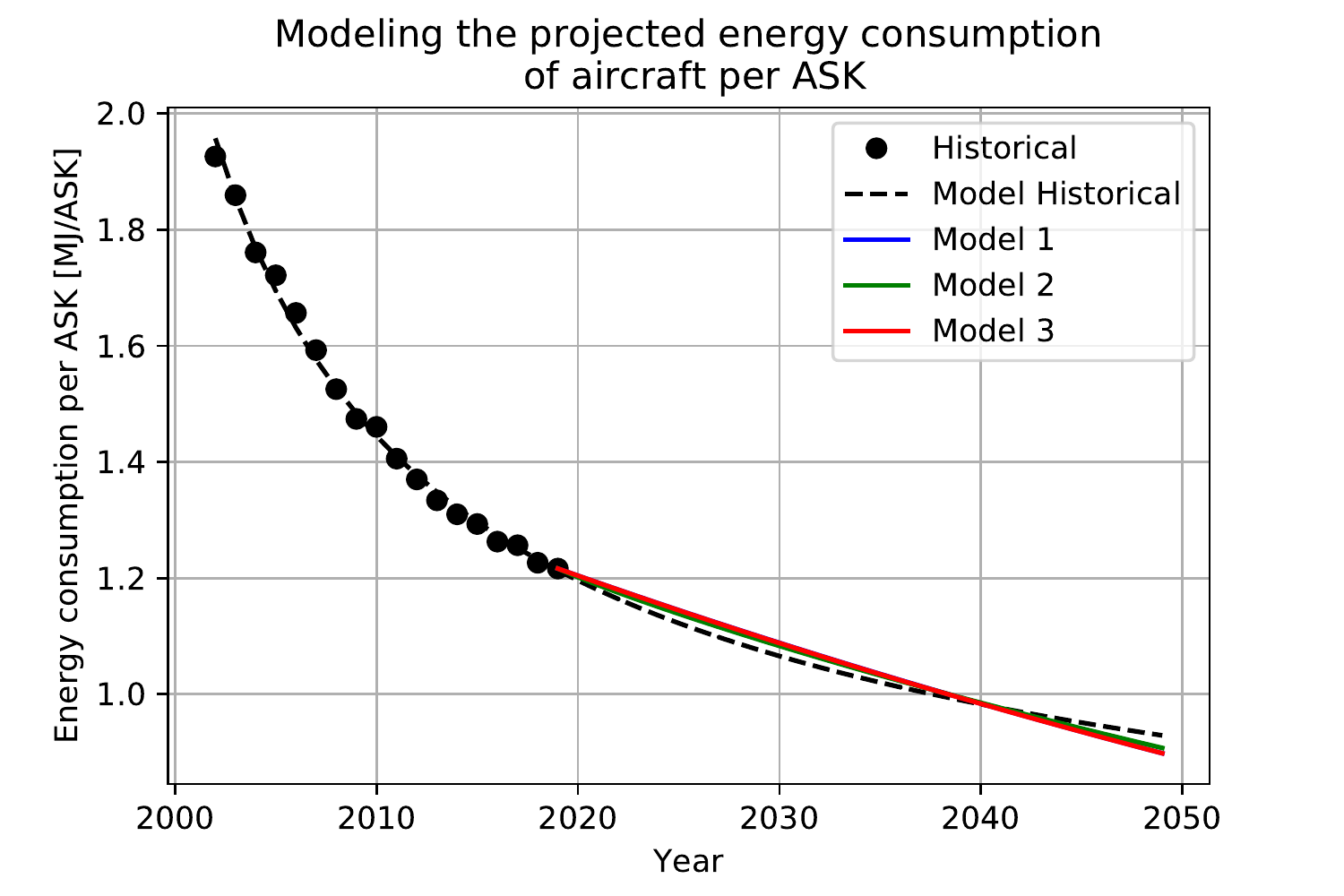}
    \caption{Models of projected aircraft energy efficiency by ASK}
    \label{fig:efficiency_proj_model}
\end{figure}

The second step consists in projecting the past trend model to obtain a trend model for future evolution. The projection of the historical model is represented by dotted lines on Figure \ref{fig:efficiency_proj_model}. In order to generate different scenarios on the evolution of this lever of action from 2020 to 2050, the modelling of this projection is carried out by considering three different models in the same way as before. Figure \ref{fig:efficiency_proj_model} shows that the optimizations of these last models give close approximations. Therefore, the simplest model of the trend efficiency by seat $Ef$, given by equation (\ref{eq:efficiency_proj}), is selected. It allows simple modelling of the trend  up to 2050 with only one coefficient $\tau$. If the trend is computed using data projected between 2020 to 2050, $\tau$ equals to 1.0\%.

\begin{equation}
\label{eq:efficiency_proj}
Ef(x) = 1.22~(1-\tau)^{x-2019}~[MJ/ASK]
\end{equation}

\begin{figure}[hbt!]
    \centering
    \includegraphics[width=1\textwidth]{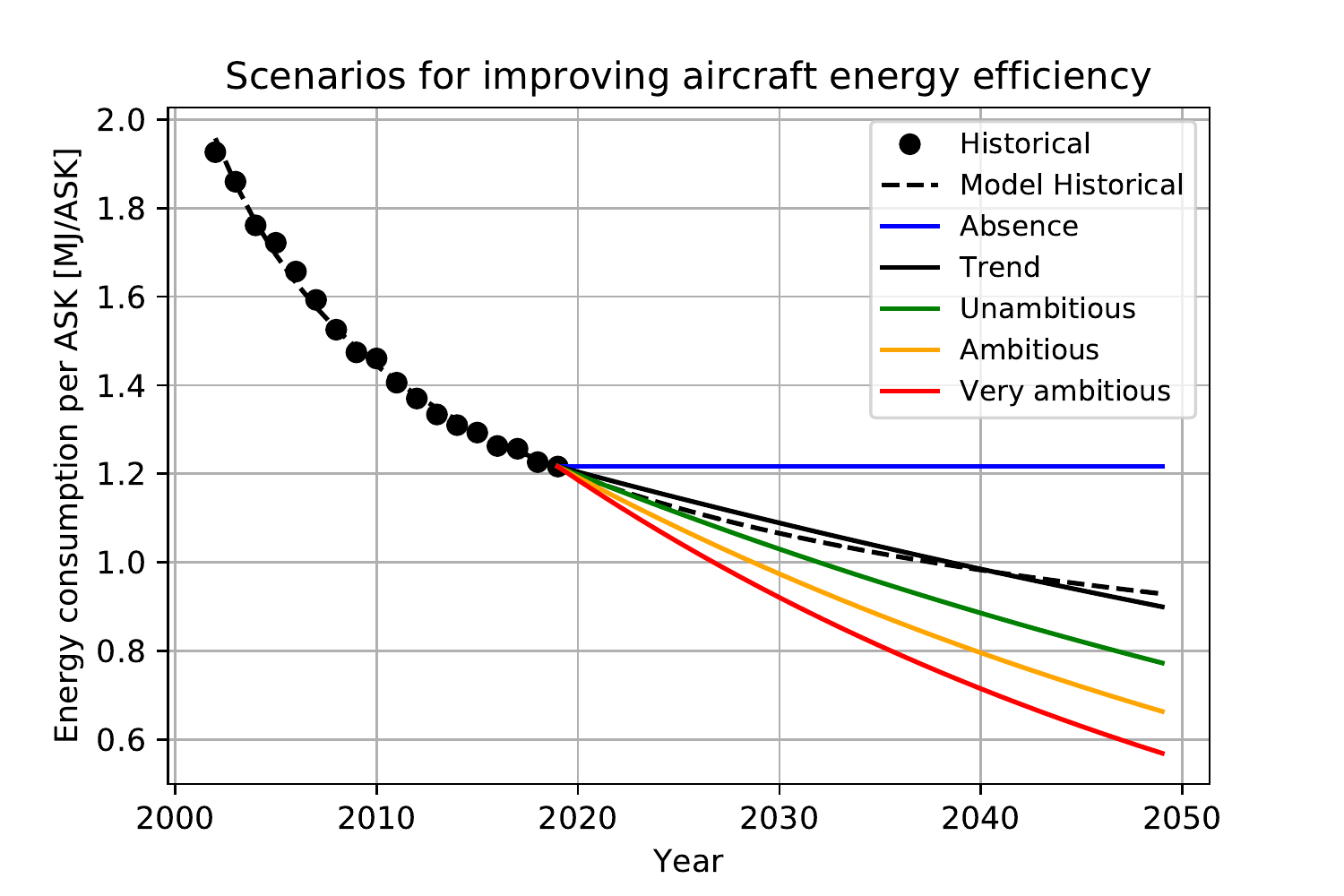}
    \caption{Scenarios for aircraft energy efficiency by ASK}
    \label{fig:efficiency_recap}
\end{figure}

Finally, the last step consists in defining different scenarios for the future by playing on the parameter $\tau$. $\tau$ equals to 0 corresponds to the "Absence" scenario in which the energy efficiency remains at the 2019 level.  The value of $\tau=1.0\%$ corresponds to the "Trend" scenario of Figure \ref{fig:efficiency_recap}. Other scenario can be studied using the model developed in step 2 and different values of $\tau$ which reflect more or less ambitious changes. The "Unambitious" scenario corresponds to a rate of 1.5\%, which corresponds to the average annual improvements over the last 5 years calculated from historical data. Similarly, the "Ambitious" and "Very ambitious" scenarios correspond respectively to a rate of 2.0\% and 2.5\%, which corresponds to the average annual improvements over the last 10 and 15 years. Figure \ref{fig:efficiency_recap} summarizes the different scenarios considered.

\subsubsection{Operations}

Energy efficiency per seat can also be improved by reducing the travelled distance, optimizing flight paths and designing better infrastructure for aircraft on the ground. All these improvements, known as flight and ground operations, represent another lever of action. These operations have been little modelled in recent decades and there is no available historical data to model their evolution. 

To  overcome the lack of data and model the evolution of operations, it is proposed to use sigmoid functions which can represent an evolution of implementation until a maximum level is reached. These model are present in many technological, sociological or economic fields \cite{kucharavy2011application, jarne2007s}. The equation (\ref{eq:sigmoid}) represents the models used in this paper.

\begin{equation}
\label{eq:sigmoid}
s(x) = \frac{V_f}{1+e^{-\alpha(x-x_0)}}
\end{equation}

\noindent
where $s$ is the sigmoid model, $x$ the year, $V_f$ the final value of the model, $\alpha$ a coefficient to set the speed of change and $x_0$ the reference year for the inflection.

In the case of operations modelling, sigmoid functions allow modelling for example the effect of specific measures to reduce consumption. The choice of the coefficients of the model allows introducing several scenarios. These scenarios, particularly the realistic one, have been established from the industrial data \cite{atag, iata}. For each scenario, it is assumed that $\alpha=0.2$ and $x_0=2030$.

\begin{itemize}
    \item Absence : no new operations are considered;
    \item Pessimistic : operational improvements are only marginally implemented and allow a reduction of consumption of 4\% compared to values in 2019, which means that $V_f=0.96$;
    \item Realistic : operational improvements are developing and allow a reduction of consumption of 8\%;
    \item Optimistic : operational improvements are widespread and allow a reduction of consumption of 12\%;
    \item Idealistic : improvements in operations are generalized and optimized and allow a reduction of consumption of 15\%.
\end{itemize}

\subsubsection{Load Factor}

To model the evolution of the aircraft load factor, an approach similar to that of efficiency is used. Indeed, historical data are available from 1991 \cite{icao} and enable trend models to be produced for describing the behaviour of the observed data. The model of the aircraft load factor, based on a sigmoid and given in equation (\ref{eq:loadfactor_hist}) as a function of the year $x$, is obtained by minimizing the RMS error between the historical data and the model. It is interesting to note that this model converges to an aircraft load factor of about 90\%. 

\begin{equation}
\label{eq:loadfactor_hist}
g(x) = 51.3 + \frac{38.7}{1+e^{-0.072(x-2000)}}~[\%]
\end{equation}

Then, to model the projections, sigmoid functions are also used. The aircraft load factor is then modelled using the equation (\ref{eq:loadfactor_proj}) with $\alpha, \beta, x_0$ coefficients. The trend model of projected data is described with coefficients $\alpha = 0.081$, $\beta = 0.15$ and $x_0 = 2030$. Different settings for these coefficients lead to the different scenarios presented in Figure \ref{eq:loadfactor_hist}. The sigmoid model allows modifying the rate of change of the aircraft load factor but one of the limits is the jump in value observed in 2020.

\begin{equation}
\label{eq:loadfactor_proj}
LF(x) = 82.4~\left(1 + \frac{\alpha}{1+e^{-\beta(x-x_0)}}\right)~[\%]
\end{equation}

\begin{figure}[hbt!]
    \centering
    \includegraphics[width=1\textwidth]{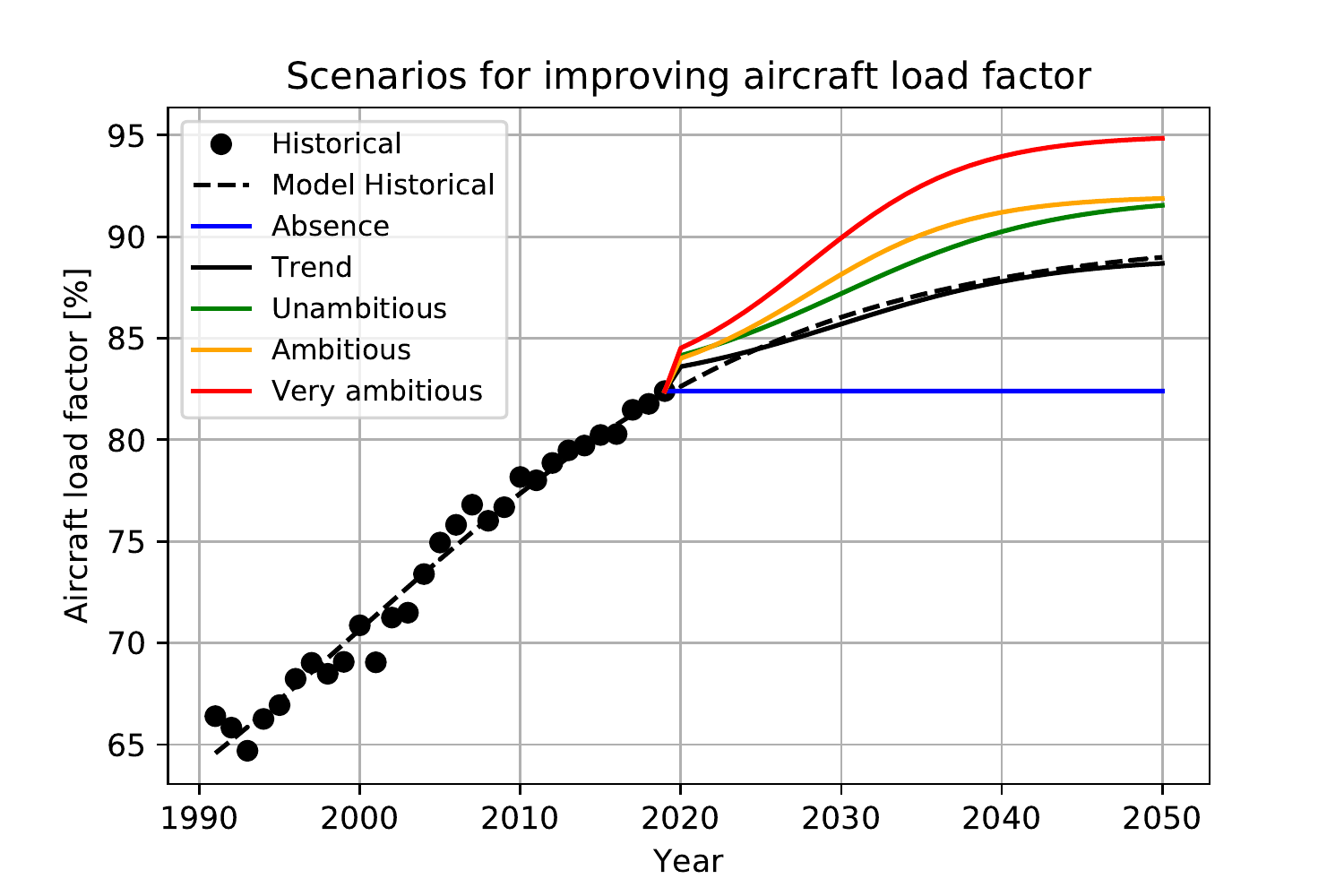}
    \caption{Scenarios for aircraft mean load factor}
    \label{fig:loadfactor_recap}
\end{figure}

\subsubsection{Energy}

One lever of action concerns the decarbonisation of energy, i.e. the reduction of the $CO_2$ content of the energy used. In the same way as for operations, this action lever is currently used marginally and modelling using sigmoid functions can be used.

To estimate the maximum decarbonisation rate of alternative fuels, an average value of about $25gCO_2/MJ$ is considered for biofuel emissions \cite{de2017life}. The results are comparable for hydrogen even if major challenges remain \cite{khandelwal2013hydrogen}. Therefore, it is considered that the decarbonisation rate of alternative fuels compared to kerosene is 70\%. The scenarios therefore focus on the proportion of the aircraft fleet that will operate on alternative fuels in the future.

For these scenarios, only the overall decarbonisation rate is modified. The latter can take values between 0\% (no aircraft has access to low-carbon fuels) and 70\% (the entire fleet has access to low-carbon fuels). The coefficients of the equation (\ref{eq:sigmoid}) are set to $\alpha=0.4$ and $x_0=2040$ to obtain trajectories consistent with the industrial data \cite{atag, iata}.

However, these scenarios will have to be refined in the future taking into account the constraints on the availability of global energy resources.

\subsubsection{Non-$CO_2$ effects}

The last major lever of action to reduce the climate impacts of aviation concerns the mitigation of non-$CO_2$ effects. In this article, only specific strategies against contrails are considered.

Many strategies to prevent the formation of contrails are being considered, both from a technological and operational point of view \cite{noppel2007overview, gierens2008review}. The technological measures mainly considered are the reduction of the quantity and size of emitted particles \cite{noppel2007overview}. From an operational point of view, modifying the flight altitude for certain atmospheric conditions is studied \cite{gierens2008review}. Quantitative studies have been performed to estimate the potential gains of these strategies. For example, different scenarios are studied in \cite{teoh2020mitigating} and lead to contrails reductions between 20\% and 91.8\%. Similar analyses are also achieved in \cite{matthes2020climate}. These different studies, valid for kerosene, must however be adapted for the use of alternative fuels. For instance, the use of hydrogen also leads to the formation of contrails, but the comparison with conventional fuels remains uncertain \cite{marquart2005upgraded, noppel2007overview}.

As with previous models, the modelling of this lever of action is based on the use of sigmoids. The scenarios considered here are extracted from \cite{teoh2020mitigating} and are given below. They are based on changes in flight altitude and the use of more efficient combustion chambers, called Dual Annular Combustor (DAC).

\begin{itemize}
    \item Absence : no strategies on contrails;
    \item Pessimistic : slight changes in altitude, which do not lead to over-consumption, are implemented on conventional engines;
    \item Realistic : more significant altitude changes, which result in slight overconsumption, are implemented on conventional engines;
    \item Optimistic : slight changes in altitude, which do not lead to over-consumption, are implemented on improved DAC engines;
    \item Idealist : more significant altitude changes, which result in slight overconsumption, are implemented on improved DAC engines.
\end{itemize}

\subsubsection{Options}

In addition to the various levers of action presented, more specific options have been included in CAST. They allow in particular studying the specific effects due to the Covid-19 epidemic as well as the impact of different economic, logistical or political measures.

To model the impact of the Covid-19 epidemic, the tool is based on projections made by IATA \cite{iata_covid}. IATA has forecast a 66\% drop in air traffic in 2020 compared to 2019 and a return of air traffic to the 2019 level by 2024. Finally, the aircraft load factor has been affected by the epidemic and IATA has forecast a load factor value of 58.5\% in 2020 against 82.4\% in 2019.

Specific measures are also implemented in CAST. For instance, economic measures to represent CORSIA agreements are also considered \cite{scheelhaase2018eu}. These are marked-based measures that compensate for emissions of $CO_2$ above the 2019 level to allow carbon-neutral growth from 2020. Secondly, another option is to model social measures concerning air transport. Indeed, 1\% of the people in the world are responsible for 50\% of $CO_2$ emissions from aviation \cite{gossling2020global}. The implemented option allows dividing by 2 the number of flights for this part of the population.

\subsection{Climate analysis}

To evaluate scenarios for aviation obtained from the models defined above, the concept of carbon budget is introduced and generalized in a simplified way to non-$CO_2$ effects in this section. The assumptions for allocating carbon budgets are also given and the analyses are carried out until 2050.

The carbon budget is an interesting concept for estimating the impact of greenhouse gases on the global average temperature \cite{matthews2009proportionality} and allows studying the ability of trajectories to reach climate targets \cite{friedlingstein2014persistent}. The latter makes it possible to relate the increase in average temperature to the cumulative quantity of $CO_2$ emissions \cite{ipcc2013ipcc}. As a consequence, it is possible to determine the remaining quota of $CO_2$ emissions to respect a maximum temperature rise. These carbon budgets yield different estimates, with significant uncertainty expressed in percentiles of Transient Climate Response to cumulative carbon Emissions (TCRE) \cite{rogelj2016differences}. Table \ref{tab:carbon_budget} summarizes the different world carbon budgets estimated by IPCC \cite{masson2018global}. To take into account Earth system feedback, $100~Gt$ must be subtracted from these budgets.

\begin{table}[!ht]
    \centering
    \caption{Remaining carbon budgets from 01.01.2018 (without Earth system feedback)}
    \vspace{10pt}
    \begin{tabular}{|c|c|c|}
        \hline
        \cellcolor{lightgray} Percentiles of TCRE & \cellcolor{lightgray} 1.5°C carbon budget & \cellcolor{lightgray} 2°C carbon budget\\
        \hline
        33\% & $840~GtCO_2$ & $2030~GtCO_2$\\
        50\% & $580~GtCO_2$ & $1500~GtCO_2$\\
        67\% & $420~GtCO_2$ & $1170~GtCO_2$\\
        \hline
    \end{tabular}
    \label{tab:carbon_budget}
\end{table}

One method to calculate the carbon budgets is to use equation (\ref{eq:carbon_budget_calculation}) extracted from \cite{rogelj2019estimating}. $CB$ represents the carbon budget, $T_{lim}$ the limit temperature rise, $T_{hist}=0.97^{\circ}C$ the temperature rise already achieved until 2015, $T_{non-CO_2}$ the impact of non-$CO_2$ effects (equals to 0.1°C for 1.5°C and to 0.2°C for 2°C), $TCRE=0.45^{\circ}C$ (for median value) and $ESF=100~Gt$ Earth system feedback.

\begin{equation}
\label{eq:carbon_budget_calculation}
CB = \frac{T_{lim}-T_{hist}-T_{non-CO_2}}{TCRE}-ESF
\end{equation}

IPCC has also taken into account the possible deployment of carbon capture and storage strategies, known as BECCS (Bio-Energy with Carbon Capture and Storage). Four scenarios have been defined in \cite{masson2018global}. P1 does not consider BECCS when P2 considers a storage capacity of $151~GtCO_2$, P3 of $414~GtCO_2$ and P4 of $1191~GtCO_2$, all by 2100.

A corrected carbon budget $CB_{c,2100}$ is defined, especially to take into account BECCS. It can be estimated with equation (\ref{eq:corrected_CB}) using the IPCC carbon budget $CB$, the correction due to Earth system feedback $ESF$, the carbon storage $BECCS$ and the $CO_2$ emissions in 2018 and 2019 $E_{CO_2,old}$.

\begin{equation}
\label{eq:corrected_CB}
CB_{c,2100} = CB + BECCS - E_{CO_2,old}
\end{equation}

Considering an analysis up to 2100, this budget is equal to the world cumulative $CO_2$ emissions between now and 2100, which gives equation (\ref{eq:sum_carbon_budget}) with $E_{CO_2,k}$ the annual world $CO_2$ emissions. 

\begin{equation}
\label{eq:sum_carbon_budget}
CB_{c,2100} = \sum_{k=2020}^{2100} E_{CO_2,k}
\end{equation}

A model with a fixed annual rate of decrease $x$ is selected to compute a reference trajectory for $CB_{c,2100}$. Equation (\ref{eq:sum_carbon_budget_ref}) is a reformulation of equation (\ref{eq:sum_carbon_budget}) with this assumption. This equation can then be solved implicitly in order to determine the annual rate of decrease $x$.

\begin{equation}
\label{eq:sum_carbon_budget_ref}
CB_{c,2100} = \sum_{k=2020}^{2100} E_{CO_2,2019}(1-x)^{k-2019} = E_{CO_2,2019} \frac{(1-x)-(1-x)^{82}}{x}
\end{equation}

To limit the analysis to 2050, $x$  being known, CAST uses equation (\ref{eq:carbon_budget_2050}) to compute the world corrected carbon budget until 2050 $CB_{c,2050}$.

\begin{equation}
\label{eq:carbon_budget_2050}
CB_{c,2050} = E_{CO_2,2019}\frac{(1-x)-(1-x)^{32}}{x}
\end{equation}

To compute the carbon budget allocated to aviation until 2050 for a target of 1.5°C or 2°C, a share of the world carbon budget must be performed. If $F$ is the rate of the carbon budget allocated to aviation, then the corrected carbon budget given to aviation until 2050 is $F.CB_{c,2050}$. $F$ is set by default in CAST to aviation's share of world $CO_2$ emissions in 2019, i.e. 2.6\%, but can be modified. Indeed, the choice of this share results from a political choice. For instance, increasing this share gives more flexibility to aviation to the detriment of other sectors, and conversely.

The approach described above is extended to non-$CO_2$ effects to compute ERF budgets. Adapting the equations for carbon budgets, a corrected equivalent carbon budget for 2100 $ECB_{c,2100}$ is estimated with equation (\ref{eq:equivalent_carbon_budget_calculation}), with $E_{GHG,old}$ the GHG emissions in 2018 and 2019 given in \cite{onu_ghg}. 

\begin{equation}
\label{eq:equivalent_carbon_budget_calculation}
ECB_{c,2100} = \frac{T_{lim}-T_{hist}}{TCRE} - ESF + BECCS - E_{GHG,old}
\end{equation}

The approach to compute the corrected equivalent carbon budget for 2050 $ECB_{c,2050}$ is then the same as before, this time considering annual GHG emissions $E_{GHG,k}$. Equation (\ref{eq:equivalent_carbon_budget_2050}) gives $ECB_{c,2050}$, to which a share $F$ must be allocated for aviation. In this case, $F$ is set by default in CAST to aviation's share of world ERF in 2011, i.e. 3.5\%.

\begin{equation}
\label{eq:equivalent_carbon_budget_2050}
ECB_{c,2050} = E_{GHG,2019}\frac{(1-x)-(1-x)^{32}}{x}
\end{equation}

Finally, this equivalent carbon budget for aviation is multiplied by the ERF coefficient for $CO_2$ in order to obtain an equivalent in ERF. This climate budget in ERF can then be compared with the adjusted ERF of aviation in 2050, which is the ERF of aviation in 2050 from which the cumulative $CO_2$ effect until 2019 is subtracted ($38.2~mW/m^2$).

\section{Results and discussions}
\label{chap:results}

In this part, CAST is used on some scenarios in order to check their compatibility with the objectives of the Paris Climate Agreements in terms of $CO_2$ emissions or ERF. First, the ATAG commitments proposed by aviation stakeholders are analysed using CAST. Then, various illustrative scenarios are developed by selecting a set of levers of action and assessed with respect to 2019 situation to highlight the potential for decreasing the climate impact of aviation.

\subsection{Analysis of ATAG commitments}

In order to detail the methodology for analyzing a scenario using CAST, a study is carried out on ATAG commitments. For the analysis, BECCS are not considered and the IPCC carbon budgets with a 50\% probability of remaining below the targeted temperature increase (1.5°C or 2°C) are taken into account.

\begin{figure}[hbt!]
    \centering
    \includegraphics[width=1\textwidth]{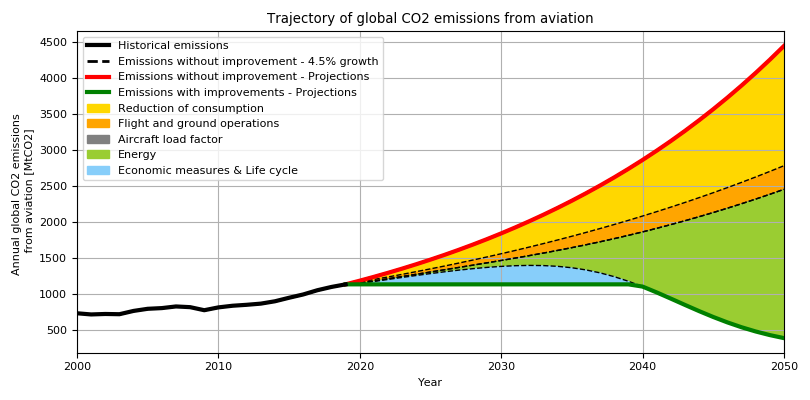}
    \caption{Modelling of 2009 ATAG commitments}
    \label{fig:scenarios_atag_2009}
\end{figure}

A modelling of 2009 ATAG commitments is shown in Figure \ref{fig:scenarios_atag_2009} which represents the trajectory of global $CO_2$ emissions for aviation. In this scenario, a 4.5\% annual growth in RPK air traffic is considered as well as a 1.5\% annual improvement in fuel efficiency (yellow part) and an optimistic improvement in operations (orange part). The evolution of the load factor is not considered and its value is therefore that of 2019. Concerning the decarbonisation of energy, using the models developed in the article, a final decarbonisation rate of 93\% for alternative fuels is necessary to obtain the trajectory defined by ATAG (green part). It is interesting to note that this value is much higher than the 70\% decarbonisation rate estimated to be achievable for biofuels or hydrogen. Lastly, to cushion the transition, economic carbon offsetting measures are being put in place to compensate for $CO_2$ emissions above the 2019 level (blue part). 

The analysis of this scenario shows that the cumulative global emissions of $CO_2$ for aviation until 2050 are equal to 30.5 Gt. In comparison, the world carbon budgets until 2050 for 1.5°C and for 2°C are respectively equal to 378 Gt and 865 Gt. Therefore, considering this scenario, aviation would consume 8.1\% of the world carbon budget for 1.5°C and 3.5\% of the world carbon budget for 2°C until 2050. Since aviation accounts for 2.6\% of global $CO_2$ emissions in 2019, it would consume more than this share in this scenario.

\begin{figure}[hbt!]
    \centering
    \includegraphics[width=1\textwidth]{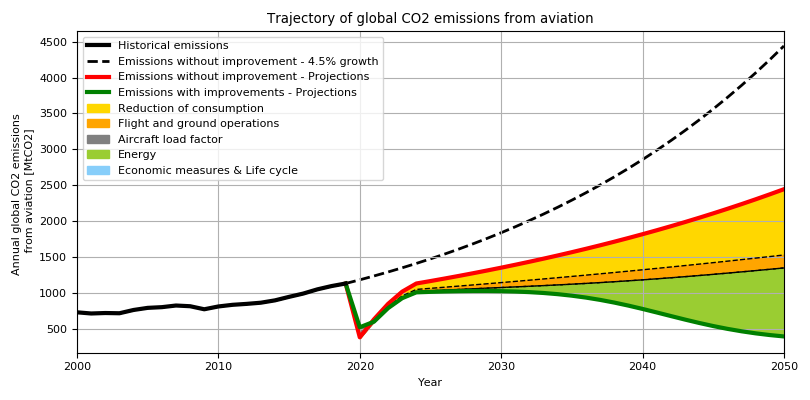}
    \caption{Modelling of 2020 ATAG commitments including Covid-19}
    \label{fig:scenarios_atag_covid}
\end{figure}

Due to Covid-19, air traffic was severely disrupted in 2020 and will be impacted for years to come. ATAG has updated its commitments to take into account the impacts of Covid-19 (Figure \ref{fig:scenarios_atag_covid}). The return of air traffic to the 2019 level is only envisaged for 2024 and the annual growth rate for the following years is estimated at 3.0\%. To model this update in CAST, the forecasts for improvements in energy efficiency and operations are kept to the 2009 commitments. The final decarbonisation rate obtained is decreased to 78\%, which is still higher than the possible expected value of 70\%.

Using the same type of analysis as for the 2009 ATAG commitments, the cumulative global emissions of $CO_2$ until 2050 are about 24.7 Gt, which corresponds to 6.5\% of the world carbon budget for 1.5°C and 2.9\% of the world carbon budget for 2°C until 2050. In the same way as for the previous scenario, aviation would consume more than the 2.6\% share in this scenario.

In terms of global ERF, 2020 ATAG commitments result in an ERF for aviation of $201.7~mW/m^2$ in 2050, i.e. an adjusted ERF of $163.5~mW/m^2$. In this scenario, aviation would consume 51.6\% and 18.9\% of the world ERF budgets in 2050 for respectively 1.5°C and for 2°C. Since aviation accounts for 3.5\% of global ERF in 2011, this scenario would consume more than this share. This large budget overshoot is due to the fact that the impact of contrails, which represents more than half of the climate impacts of aviation, is not mitigated in the ATAG commitments. This result shows the importance of integrating measures against contrails in the future.

\subsection{Simulation and analysis of three illustrative scenarios}

The objective of this part is to use CAST to simulate and analyse some illustrative scenarios. For all these case studies, BECCS are not considered and the IPCC carbon budgets with a 50\% probability of remaining below the temperature target are taken into account. The studies carried out for these scenarios are limited to fixed allocated shares for aviation that correspond to current impacts, i.e. 2.6\% for global $CO_2$ emissions and 3.5\% for global ERF.

\subsubsection{Presentation of illustrative scenarios}

Three illustrative scenarios are defined according to different levels of technological development. The settings for these scenarios are based on the models for the levers of action in Section \ref{chap:models}.

\begin{enumerate}
    \item \textit{Trend scenario for aircraft efficiency and load factor considering a kerosene-fuelled fleet without new operation}: Trend scenarios are considered for the evolution of the aircraft energy consumption (1\% annual improvement) and load factor. Improvements in operations are not considered. Moreover, it is assumed that only kerosene continues to be used as aircraft fuel. Using these assumptions, the global $CO_2$ emissions per RPK would be $89~gCO_2/RPK$ in 2050.
    
    \item \textit{Trend scenario for aircraft efficiency and load factor including low-carbon fuels and new operations}: Trend scenarios are considered for the evolution of the aircraft energy consumption (1\% annual improvement) and load factor. For operations, a realistic improvement is taken into account, in accordance with the models in the previous section. Moreover, a transition to low-carbon fuels (70\% reduction compared to kerosene) for half of the fleet by 2050 is considered. This corresponds in the models to a total energy decarbonisation for the entire fleet of 35\%. Using these assumptions, the global $CO_2$ emissions per RPK would be $54~gCO_2/RPK$ in 2050.
    
    \item \textit{Technology-based scenario}: Technologies are pushed forward with optimistic assumptions. First, the annual rate of improvement in aircraft fuel efficiency is 1.5\%, which corresponds to the average value for the last 5 years. In comparison, ATAG also considers annual improvements of 1.5\%. Next, it is assumed that the entire fleet will be able to be fuelled by alternative low-carbon fuels (70\% reduction compared to kerosene) by 2050. Using these assumptions, the global $CO_2$ emissions per RPK would be $20~gCO_2/RPK$ in 2050.
\end{enumerate}

The level of air traffic, modelled with the annual growth rate of RPK, is considered variable in these scenarios. Four distinct cases are studied: estimated trend of traffic growth before Covid-19 (4.5\%), estimated trend of traffic growth after Covid-19 (3\%), stagnation of traffic at the 2019 level and traffic necessary to equal the carbon budget for 2°C. The effects of Covid-19 are included in the last three cases for the level of traffic.

\subsubsection{Analysis for $CO_2$ emissions}

In this section, illustrative scenarios are analysed in terms of $CO_2$ emissions and carbon budgets.

Firstly, the analysis of the trend scenario excluding low-carbon energy with the estimated growth of air traffic before Covid-19 shows cumulative $CO_2$ emissions of 60.0 Gt. It largely exceeds the carbon budgets for 1.5°C and 2°C allocated to aviation, which are respectively 10.0 Gt and 22.8 Gt. These cumulative $CO_2$ emissions corresponds to 6.9\% of the 2°C world carbon budget for 2050. Similarly, considering the projections after the Covid-19 crisis, the cumulative $CO_2$ emissions are equal to 38.8 Gt, which also exceeds the carbon budgets allocated to aviation. The results are similar for the two other scenarios. Indeed, the trend scenario including low-carbon energy leads to cumulative $CO_2$ emissions of 47.9 Gt for a RPK growth of 4.5\% and to 31.5 Gt for a RPK growth of 3\%, and the technology-based scenario leads to cumulative $CO_2$ emissions of 27.0 Gt for a RPK growth of 4.5\% and to 23.4 Gt for a RPK growth of 3\%. Therefore, in order to respect a trajectory compatible with the Paris Climate Agreements for these scenarios considering these allocated carbon budgets, air traffic growth projections must be reduced.

\begin{figure}[hbt!]
    \centering
    \includegraphics[width=1\textwidth]{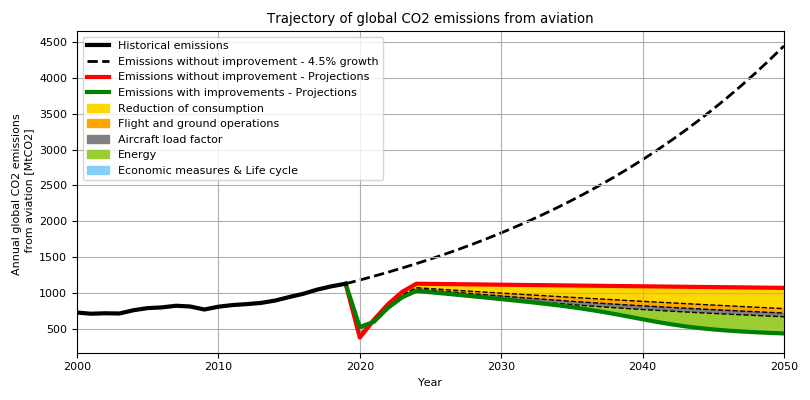}
    \caption{Trend scenario including low-carbon energy which respects 2°C carbon budget}
    \label{fig:scenarios_example_co2}
\end{figure}

The analysis is then performed for air traffic that remains at the 2019 level. In this case, the cumulative $CO_2$ emissions amount to 27.2 Gt for the trend scenario excluding low-carbon energy and to 23.0 Gt for the trend scenario including low-carbon energy. The carbon budgets are therefore exceeded for a stagnation of air traffic. Thus, to comply with the Paris Climate Agreements with these assumptions, it is therefore necessary to reduce air traffic compared to 2019. In order to respect the carbon budget for 2°C, air traffic must be reduced by 1.8\% per year in the trend scenario excluding low-carbon energy and by 0.2\% in the trend scenario including low-carbon energy. For instance, Figure \ref{fig:scenarios_example_co2} represents the aviation carbon trajectory in the case of the trend scenario including low-carbon energy.

Finally, for the technology-based scenario, a reduction in air traffic is not necessary. Indeed, the cumulative $CO_2$ emissions of this scenario amount to 18.2 Gt when considering the air traffic at the level of 2019. As a consequence, aviation would only consume 2.1\% of the global 2°C carbon budget, which respects the carbon budget allocated to aviation for 2°C. A traffic growth of 2.6\% compared to 2019 is even possible while respecting the 2°C carbon budget. However, if the assessment of this scenario is performed compared to the carbon budget allocated for 1.5°C, a decrease in air traffic is necessary as for trend scenarios.

Table \ref{tab:results} summarises the main results. Illustrative scenarios with trend RPK growth are not compatible with carbon trajectories corresponding to the Paris climate agreements with an allocated share of 2.6\%. For 2°C, a decrease in air traffic is necessary considering the trend scenarios whereas a RPK growth is possible in the technology-based scenario. For 1.5°C, all illustrative scenarios lead to a drastic air traffic decrease.

\begin{table}[!ht]
    \centering
    \caption{Results for the analysis of illustrative scenarios in terms of carbon budgets}
    \begin{tabular}{|c|c|c|c|}
            \cline{2-4}
    \multicolumn{1}{c|}{} & 
        \cellcolor{lightgray} \parbox{0.2\textwidth}{\centering \vspace{1.0ex}Illustrative scenario  1\vspace{1.0ex}} &
        \cellcolor{lightgray} \parbox{0.2\textwidth}{\centering \vspace{1.0ex}Illustrative scenario  2\vspace{1.0ex}} & \cellcolor{lightgray} \parbox{0.2\textwidth}{\centering \vspace{1.0ex}Illustrative scenario  3\vspace{1.0ex}}\\
        \hline
        \parbox{0.2\textwidth}{\centering \vspace{1.0ex}Scenario characteristics\vspace{1.0ex}} & \parbox{0.2\textwidth}{\centering \vspace{1.0ex}Trend scenario for the energy efficiency (-1\% per year) excluding low-carbon energy\vspace{1.0ex}} & 
        \parbox{0.2\textwidth}{\centering \vspace{1.0ex}Trend scenario for the energy efficiency (-1\% per year) including low-carbon energy for half of the fleet (-35\% in 2050)\vspace{1.0ex}}  & \parbox{0.2\textwidth}{\centering \vspace{1.0ex}Technology-based scenario for energy efficiency (-1.5\% per year) including low-carbon energy for all the fleet (-70\% in 2050)\vspace{1.0ex}}\\
        \hline
        \parbox{0.2\textwidth}{\centering \vspace{1.0ex}Share of the 2°C world carbon budget consumed for a 3\% growth rate\vspace{1.0ex}} & 
        \parbox{0.2\textwidth}{\centering \vspace{1.0ex}10.3\%\vspace{1.0ex}} & 
        \parbox{0.2\textwidth}{\centering \vspace{1.0ex}8.3\%\vspace{1.0ex}} & 
        \parbox{0.2\textwidth}{\centering \vspace{1.0ex}6.2\%\vspace{1.0ex}}\\
        \hline
        \parbox{0.2\textwidth}{\centering \vspace{1.0ex}Air traffic growth rate to respect a 2.6\% share for aviation for 2°C\vspace{1.0ex}} &
        \parbox{0.2\textwidth}{\centering \vspace{1.0ex}-1.8\%\vspace{1.0ex}} & 
        \parbox{0.2\textwidth}{\centering \vspace{1.0ex}-0.2\%\vspace{1.0ex}}  & 
        \parbox{0.2\textwidth}{\centering \vspace{1.0ex}2.6\%\vspace{1.0ex}}\\
        \hline
    \end{tabular}
    \label{tab:results}
\end{table}

\begin{figure}[hbt!]
    \centering
    \includegraphics[width=0.85\textwidth]{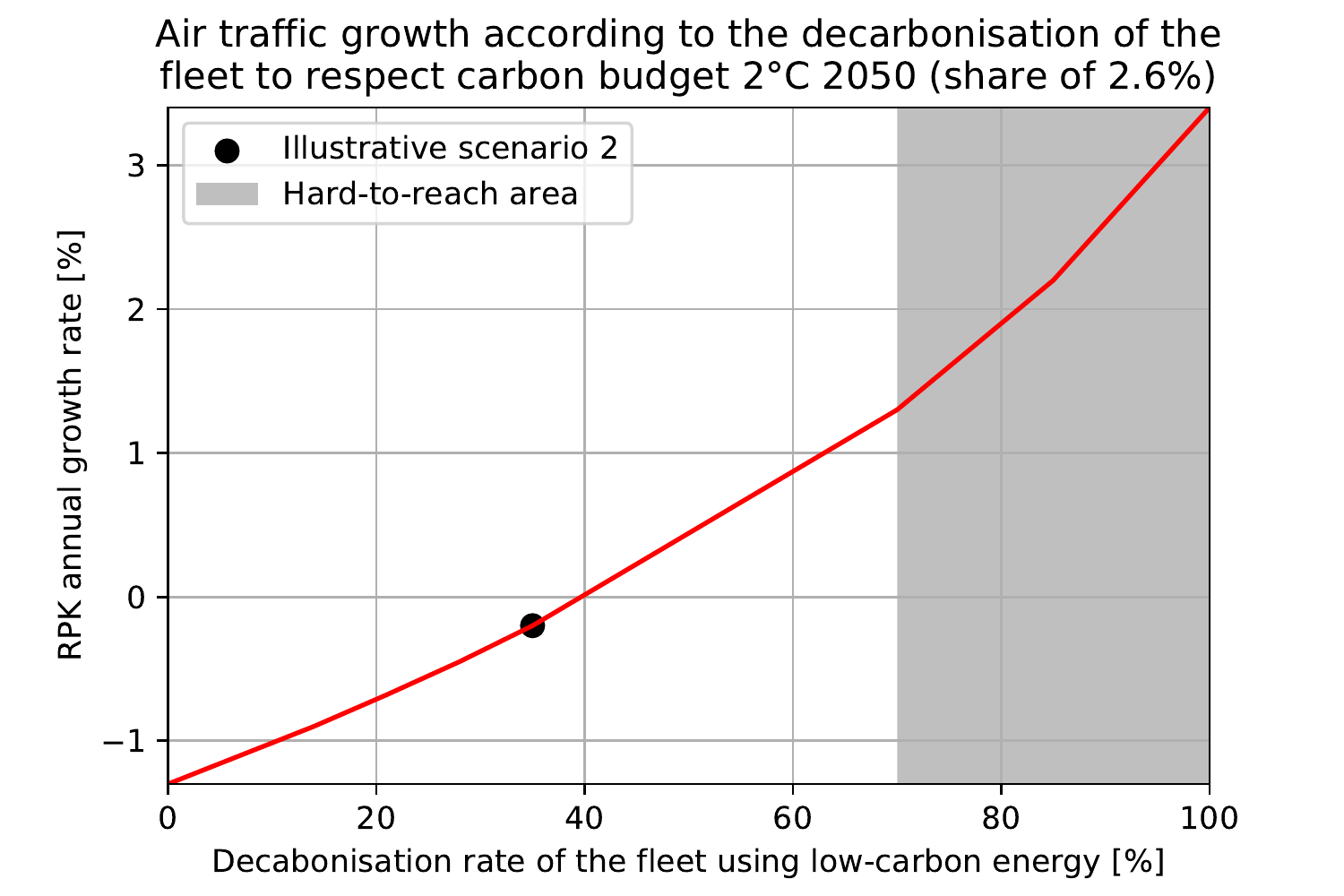}
    \caption{Influence of the decarbonisation rate on trend scenario including operations}
    \label{fig:decarbonisation}
\end{figure}

Another analysis can be conducted using CAST. Indeed, given the high uncertainty on the availability of energy resources, different studies can be carried out with the global decarbonisation rate of the fleet in 2050 as a variable. Figure \ref{fig:decarbonisation} represents for example the possible RPK growth rate as a function of the decarbonisation rate of the fleet for carbon budgets at 2°C and an allocated share of 2.6\% on illustrative scenario 2 (trend scenario including low-carbon energy). The area in the figure that requires alternative fuel with a decarbonisation rate of more than 70\% compared to kerosene, which corresponds to the hypothesis of maximum achievable decarbonisation rate, is highlighted in grey.

\subsubsection{Analysis for ERF}

In this part, illustrative scenarios are analysed in terms of ERF.

Analyses with CAST show that no illustrative scenario is compatible with climate budgets in ERF without a strong decrease in air traffic. To limit the decrease, strategies against contrails can be put in place. To mitigate contrails, widespread altitude changes with more efficient combustion chambers, described in Section \ref{chap:models}, are considered.

\begin{figure}[hbt!]
    \centering
    \includegraphics[width=1\textwidth]{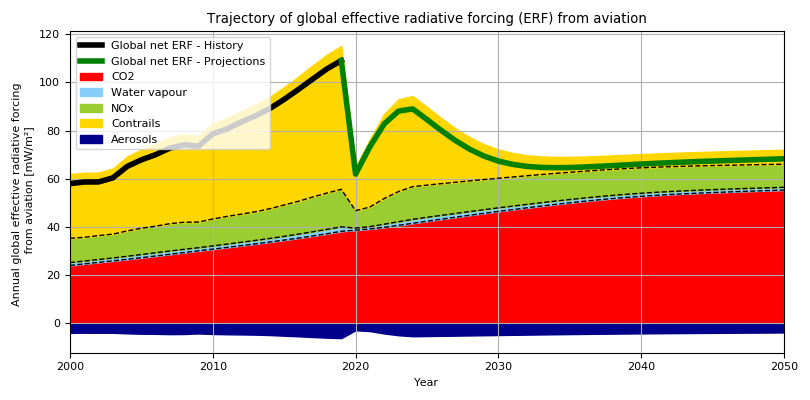}
    \caption{Technology-based scenario which respects 2°C ERF budget}
    \label{fig:scenarios_example_erf}
\end{figure}

First, the technology-based scenario including strategies against contrails is analysed in terms of climate budget in ERF. To limit the temperature increase to 2°C for a 3.5\% share of the climate budget allocated to aviation, the climate budget is respected with an annual growth rate of 0.8\%, which results in an adjusted ERF from aviation of $30.1~mW/m^2$ in 2050, i.e. a global ERF from aviation of $68.3~mW/m^2$ in 2050. Figure \ref{fig:scenarios_example_erf} represents the trajectory in ERF for this scenario. Equivalent analyses for the other illustrative scenarios are performed and lead to an RPK annual decrease of 1.3\% for the trend scenario excluding low-carbon energy and an annual decrease of 0.5\% for the trend scenario including low-carbon energy. However, when it comes to limiting the temperature increase to 1.5°C, whatever the illustrative scenario, a strong decrease in air traffic is necessary as for carbon budget analyses.

The analysis shows that ERF trajectories give different traffic levels than for carbon trajectories for 2°C. However, because of uncertainties and methodology, the use of carbon trajectories is recommended.

\section{Conclusions and future work}
\label{chap:conclusions}

In this paper, the methodology and models to develop the tool CAST for simulating and assessing climatic scenarios for the aviation industry are presented. This tool is used to simulate scenarios concerning the future climate impacts of aviation, and to assess their compatibility with the Paris climate agreements.

Regarding the methodology and the models, two main themes are addressed. Firstly, the evolution of aviation is modelled via different levers of action, like the level of air traffic, the fuel consumption efficiency or the use of low-carbon fuel, that are linked via an adapted Kaya type equation. To model these different levers of action, several strategies are used. For those with historical data, deterministic models are developed to define trend scenarios. For the others, hypotheses from the scientific literature are taken into account and projections are made. Secondly, climate models are used both to estimate the climate impact of aviation, but also to assess the compatibility of the trajectories with the Paris climate agreements. In addition to $CO_2$ emissions, non-$CO_2$ effects are considered using aggregated models from the scientific literature to estimate the impacts in terms of ERF. The evaluation of the scenarios is based on the notion of carbon budgets.

As examples, several scenarios are assessed with CAST. First of all, ATAG commitments are modelled and compared with trajectories compatible with the Paris Climate Agreements. The most recent ATAG commitments would result in a consumption of 3.0\% (respectively 6.8\%) of the world carbon budget for limiting the temperature increase to 2°C (respectively 1.5°C). This represents more than the 2.6\% share of global $CO_2$ emissions from aviation in 2019. Note that, in these commitments, the non-$CO_2$ effects are not taken into account, even though they currently account for about 2/3 of the global ERF of aviation. Then, different scenarios are simulated to take into account different levels of technological improvements. Regarding the compatibility of these scenarios with the Paris climate agreements for 2°C for $CO_2$ emissions and considering a 2.6\% share for the allocated carbon budget, the evolution of world air traffic is expected to be between an annual traffic decrease of 1.8\% (trend scenario without new fuels) and an annual growth of 2.6\% (ambitious scenario including low carbon fuels). However, air traffic should decrease drastically to be compatible with a +1.5°C trajectory. Lastly, additional studies on the non-$CO_2$ effects show the importance of implementing specific strategies to refine the possible scenarios for aviation.

Although CAST is already a mature tool for simulating and assessing climate scenarios, some limitations remain to fully analyse scenarios. First, regarding the decarbonisation of alternative fuels, constraints on the availability of energy resources (land available for biofuels, low-carbon electricity available for hydrogen production) are not addressed. These aspects will be taken into account in a future version of CAST. Second, some models represent the future evolution in a simplified way. For instance, the different scenarios considered for the evolution of the different levers of action are projected models taking into account current trends and knowledge. A better link between these projections and the envisaged future technologies will be implemented in a future version of CAST using a bottom-up approach. This would allow more accurate modelling of technologies and fleet renewal impacts. Subsequently, to improve approaches for non-$CO_2$ effects, methodologies using Global Warming Potential (GWP) indicators are envisaged, such as the modelling of other strategies to mitigate non-$CO_2$ effects. Finally, for most of the studied scenarios, climate constraints are based on an allocated share corresponding to the current impacts of aviation. This share could be determined by coupling these studies with social-economic parameters in order to make trade-offs regarding the distribution of carbon budgets.

%

\section*{Acknowledgements}

The authors would like to thank all the people who took part in CAST beta testing for their relevant feedback. This study is supported by ISAE-SUPAERO, within the framework of the research chair CEDAR (Chair for Eco-Design of AiRcraft).

\section*{Supplementary material}

CAST is available at: http://cast.isae-supaero.fr/






\bibliographystyle{elsarticle-num}
\bibliography{bibliography}

\begin{thebibliography}{10}
\expandafter\ifx\csname url\endcsname\relax
  \def\url#1{\texttt{#1}}\fi
\expandafter\ifx\csname urlprefix\endcsname\relax\def\urlprefix{URL }\fi
\expandafter\ifx\csname href\endcsname\relax
  \def\href#1#2{#2} \def\path#1{#1}\fi

\bibitem{stocker2013climate}
T.~F. Stocker, D.~Qin, G.-K. Plattner, M.~Tignor, S.~K. Allen, J.~Boschung,
  A.~Nauels, Y.~Xia, V.~Bex, P.~M. Midgley, et~al., Climate change 2013: The
  physical science basis, Contribution of working group I to the fifth
  assessment report of the intergovernmental panel on climate change 1535.

\bibitem{ipcc2007physical}
C.~C. IPCC, The physical science basis (2007).

\bibitem{ipcc2013ipcc}
W.~IPCC, Ipcc, 2013: summary for policymakers, Climate change.

\bibitem{schleussner2016science}
C.-F. Schleussner, J.~Rogelj, M.~Schaeffer, T.~Lissner, R.~Licker, E.~M.
  Fischer, R.~Knutti, A.~Levermann, K.~Frieler, W.~Hare, Science and policy
  characteristics of the paris agreement temperature goal, Nature Climate
  Change 6~(9) (2016) 827--835.

\bibitem{masson2018global}
V.~Masson-Delmotte, P.~Zhai, H.-O. P{\"o}rtner, D.~Roberts, J.~Skea, P.~R.
  Shukla, A.~Pirani, W.~Moufouma-Okia, C.~P{\'e}an, R.~Pidcock, et~al., Global
  warming of 1.5 c, An IPCC Special Report on the impacts of global warming of
  1.

\bibitem{sterman2012climate}
J.~Sterman, T.~Fiddaman, T.~R. Franck, A.~Jones, S.~McCauley, P.~Rice,
  E.~Sawin, L.~Siegel, Climate interactive: the c-roads climate policy model,
  System Dynamics Review.

\bibitem{strapasson2020modelling}
A.~Strapasson, J.~Woods, V.~P{\'e}rez-Cirera, A.~Elizondo, D.~Cruz-Cano,
  J.~Pestiaux, M.~Cornet, R.~Chaturvedi, Modelling carbon mitigation pathways
  by 2050: Insights from the global calculator, Energy Strategy Reviews (2020)
  100494.

\bibitem{spielmann2008environmental}
M.~Spielmann, P.~de~Haan, R.~W. Scholz, Environmental rebound effects of
  high-speed transport technologies: a case study of climate change rebound
  effects of a future underground maglev train system, Journal of Cleaner
  Production 16~(13) (2008) 1388--1398.

\bibitem{bigo2020transports}
A.~Bigo, Les transports face au d{\'e}fi de la transition {\'e}nerg{\'e}tique.
  explorations entre pass{\'e} et avenir, technologie et sobri{\'e}t{\'e},
  acc{\'e}l{\'e}ration et ralentissement., Ph.D. thesis, Institut Polytechnique
  de Paris (2020).

\bibitem{cantarero2019decarbonizing}
M.~M.~V. Cantarero, Decarbonizing the transport sector: The promethean
  responsibility of nicaragua, Journal of environmental management 245 (2019)
  311--321.

\bibitem{zhang2020targeting}
L.~Zhang, Z.~Li, X.~Jia, R.~R. Tan, F.~Wang, Targeting carbon emissions
  mitigation in the transport sector--a case study in urumqi, china, Journal of
  Cleaner Production (2020) 120811.

\bibitem{lee2009aviation}
D.~S. Lee, D.~W. Fahey, P.~M. Forster, P.~J. Newton, R.~C. Wit, L.~L. Lim,
  B.~Owen, R.~Sausen, Aviation and global climate change in the 21st century,
  Atmospheric Environment 43~(22-23) (2009) 3520--3537.

\bibitem{ramaswamy2019radiative}
V.~Ramaswamy, W.~Collins, J.~Haywood, J.~Lean, N.~Mahowald, G.~Myhre, V.~Naik,
  K.~Shine, B.~Soden, G.~Stenchikov, et~al., Radiative forcing of climate: the
  historical evolution of the radiative forcing concept, the forcing agents and
  their quantification, and applications, Meteorological Monographs 59 (2019)
  14--1.

\bibitem{lee2020contribution}
D.~Lee, D.~Fahey, A.~Skowron, M.~Allen, U.~Burkhardt, Q.~Chen, S.~Doherty,
  S.~Freeman, P.~Forster, J.~Fuglestvedt, et~al., The contribution of global
  aviation to anthropogenic climate forcing for 2000 to 2018, Atmospheric
  Environment 244 (2020) 117834.

\bibitem{grewe2017mitigating}
V.~Grewe, K.~Dahlmann, J.~Flink, C.~Fr{\"o}mming, R.~Ghosh, K.~Gierens,
  R.~Heller, J.~Hendricks, P.~J{\"o}ckel, S.~Kaufmann, et~al., Mitigating the
  climate impact from aviation: achievements and results of the dlr wecare
  project, Aerospace 4~(3) (2017) 34.

\bibitem{karcher2018formation}
B.~K{\"a}rcher, Formation and radiative forcing of contrail cirrus, Nature
  communications 9~(1) (2018) 1--17.

\bibitem{atag}
A.~T.~A. Group, \href{https://www.atag.org/}{World data on aviation},
  https://www.atag.org/ (2020).
\newline\urlprefix\url{https://www.atag.org/}

\bibitem{cames2015emission}
M.~Cames, J.~Graichen, A.~Siemons, V.~Cook, Emission reduction targets for
  international aviation and shipping, Directorate General for Internal
  Policies; European Parliament—Policy Department A; Economic and Scientific
  Policy: Bruxelles, Belgium.

\bibitem{linke2020glowopt}
F.~Linke, K.~Radhakrishnan, V.~Grewe, R.~Vos, M.~Nikla{\ss}, B.~L{\"u}hrs,
  F.~Yin, I.~Dedoussi, Glowopt-a new approach towards global-warming-optimized
  aircraft design, in: 1st Aerospace Europe Conference (AEC), 2020.

\bibitem{johanning2014first}
A.~Johanning, D.~Scholz, A first step towards the integration of life cycle
  assessment into conceptual aircraft design, Deutsche Gesellschaft f{\"u}r
  Luft-und Raumfahrt-Lilienthal-Oberth eV, 2014.

\bibitem{pinheiro2020sustainability}
S.~Pinheiro~Melo, A.~Barke, F.~Cerdas, C.~Thies, M.~Mennenga, T.~S. Spengler,
  C.~Herrmann, Sustainability assessment and engineering of emerging aircraft
  technologies—challenges, methods and tools, Sustainability 12~(14) (2020)
  5663.

\bibitem{kurniawan2011comparison}
J.~S. Kurniawan, S.~Khardi, Comparison of methodologies estimating emissions of
  aircraft pollutants, environmental impact assessment around airports,
  Environmental Impact Assessment Review 31~(3) (2011) 240--252.

\bibitem{ribeiro2020environmental}
J.~Ribeiro, F.~Afonso, I.~Ribeiro, B.~Ferreira, H.~Policarpo, P.~Pe{\c{c}}as,
  F.~Lau, Environmental assessment of hybrid-electric propulsion in conceptual
  aircraft design, Journal of Cleaner Production 247 (2020) 119477.

\bibitem{de2017life}
S.~De~Jong, K.~Antonissen, R.~Hoefnagels, L.~Lonza, M.~Wang, A.~Faaij,
  M.~Junginger, Life-cycle analysis of greenhouse gas emissions from renewable
  jet fuel production, Biotechnology for biofuels 10~(1) (2017) 64.

\bibitem{zhang2020prospects}
H.~Zhang, Y.~Fang, M.~Wang, L.~Appels, Y.~Deng, Prospects and perspectives
  foster enhanced research on bio-aviation fuels, Journal of Environmental
  Management 274 (2020) 111214.

\bibitem{yilmaz2012investigation}
{\.I}.~Y{\i}lmaz, M.~{\.I}lba{\c{s}}, M.~Ta{\c{s}}tan, C.~Tarhan, Investigation
  of hydrogen usage in aviation industry, Energy conversion and management 63
  (2012) 63--69.

\bibitem{aakerman2005sustainable}
J.~{\AA}kerman, Sustainable air transport----on track in 2050, Transportation
  Research Part D: Transport and Environment 10~(2) (2005) 111--126.

\bibitem{terrenoire2019contribution}
E.~Terrenoire, D.~Hauglustaine, T.~Gasser, O.~Penanhoat, The contribution of
  carbon dioxide emissions from the aviation sector to future climate change,
  Environmental research letters 14~(8) (2019) 084019.

\bibitem{sharmina2020decarbonising}
M.~Sharmina, O.~Edelenbosch, C.~Wilson, R.~Freeman, D.~Gernaat, P.~Gilbert,
  A.~Larkin, E.~Littleton, M.~Traut, D.~Van~Vuuren, et~al., Decarbonising the
  critical sectors of aviation, shipping, road freight and industry to limit
  warming to 1.5--2° c, Climate Policy (2020) 1--20.

\bibitem{qiu2017carbon}
R.~Qiu, J.~Xu, Z.~Zeng, Carbon emission allowance allocation with a mixed
  mechanism in air passenger transport, Journal of environmental management 200
  (2017) 204--216.

\bibitem{scheelhaase2018eu}
J.~Scheelhaase, S.~Maertens, W.~Grimme, M.~Jung, Eu ets versus corsia--a
  critical assessment of two approaches to limit air transport's co2 emissions
  by market-based measures, Journal of Air Transport Management 67 (2018)
  55--62.

\bibitem{icao}
I.~C.~A. Organization,
  \href{https://www.icao.int/about-icao/Pages/annual-reports.aspx}{World data
  on aviation}, https://www.icao.int/about-icao/Pages/annual-reports.aspx
  (2020).
\newline\urlprefix\url{https://www.icao.int/about-icao/Pages/annual-reports.aspx}

\bibitem{gossling2020global}
S.~G{\"o}ssling, A.~Humpe, The global scale, distribution and growth of
  aviation: Implications for climate change, Global Environmental Change 65
  (2020) 102194.

\bibitem{iea}
I.~E. Agency, \href{https://www.iea.org/sankey/}{World data on energy},
  https://www.iea.org/sankey/ (2020).
\newline\urlprefix\url{https://www.iea.org/sankey/}

\bibitem{ademe}
ADEME,
  \href{https://www.bilans-ges.ademe.fr/docutheque/docs/%5BBase%20Carbone%5D%20Documentation%20g%C3%A9n%C3%A9rale%20v11.5.pdf}{French
  and european data for global greenhouse gas emissions},
  https://www.bilans-ges.ademe.fr/docutheque/docs/
  (2020).
\newline\urlprefix\url{https://www.bilans-ges.ademe.fr/docutheque/docs/%5BBase%20Carbone%5D%20Documentation%20g%C3%A9n%C3%A9rale%20v11.5.pdf}

\bibitem{stratton2011impact}
R.~W. Stratton, P.~J. Wolfe, J.~I. Hileman, Impact of aviation non-co2
  combustion effects on the environmental feasibility of alternative jet fuels,
  Environmental science \& technology 45~(24) (2011) 10736--10743.

\bibitem{globalcarbon}
G.~C. Project, \href{https://www.globalcarbonproject.org/}{World data for
  global greenhouse gas emissions}, https://www.globalcarbonproject.org/
  (2020).
\newline\urlprefix\url{https://www.globalcarbonproject.org/}

\bibitem{mckinney-proc-scipy-2010}
{W}es {M}c{K}inney, {D}ata {S}tructures for {S}tatistical {C}omputing in
  {P}ython, in: {S}t\'efan van~der {W}alt, {J}arrod {M}illman (Eds.),
  {P}roceedings of the 9th {P}ython in {S}cience {C}onference, 2010, pp. 56 --
  61.
\newblock \href {http://dx.doi.org/10.25080/Majora-92bf1922-00a}
  {\path{doi:10.25080/Majora-92bf1922-00a}}.

\bibitem{2020SciPy-NMeth}
P.~Virtanen, R.~Gommers, T.~E. Oliphant, M.~Haberland, T.~Reddy, D.~Cournapeau,
  E.~Burovski, P.~Peterson, W.~Weckesser, J.~Bright, S.~J. {van der Walt},
  M.~Brett, J.~Wilson, K.~J. Millman, N.~Mayorov, A.~R.~J. Nelson, E.~Jones,
  R.~Kern, E.~Larson, C.~J. Carey, {\.I}.~Polat, Y.~Feng, E.~W. Moore,
  J.~{VanderPlas}, D.~Laxalde, J.~Perktold, R.~Cimrman, I.~Henriksen, E.~A.
  Quintero, C.~R. Harris, A.~M. Archibald, A.~H. Ribeiro, F.~Pedregosa, P.~{van
  Mulbregt}, {SciPy 1.0 Contributors}, {{SciPy} 1.0: Fundamental Algorithms for
  Scientific Computing in Python}, Nature Methods 17 (2020) 261--272.
\newblock \href {http://dx.doi.org/10.1038/s41592-019-0686-2}
  {\path{doi:10.1038/s41592-019-0686-2}}.

\bibitem{ipywidgets}
JupyterDevTeam,
  \href{https://github.com/jupyter-widgets/ipywidgets}{ipywidgets: Interactive
  html widgets}, https://github.com/jupyter-widgets/ipywidgets (2020).
\newline\urlprefix\url{https://github.com/jupyter-widgets/ipywidgets}

\bibitem{ipympl}
JupyterDevTeam, \href{https://github.com/matplotlib/ipympl}{ipympl: Matplotlib
  jupyter integration}, https://github.com/matplotlib/ipympl (2020).
\newline\urlprefix\url{https://github.com/matplotlib/ipympl}

\bibitem{voila}
Quantstack, \href{https://github.com/voila-dashboards/voila}{voila: Voilà
  turns jupyter notebooks into standalone web applications},
  https://github.com/voila-dashboards/voila (2020).
\newline\urlprefix\url{https://github.com/voila-dashboards/voila}

\bibitem{kaya1997environment}
Y.~Kaya, K.~Yokobori, et~al., Environment, energy, and economy: strategies for
  sustainability, United Nations University Press Tokyo, 1997.

\bibitem{friedl2003determinants}
B.~Friedl, M.~Getzner, Determinants of co2 emissions in a small open economy,
  Ecological economics 45~(1) (2003) 133--148.

\bibitem{schandl2016decoupling}
H.~Schandl, S.~Hatfield-Dodds, T.~Wiedmann, A.~Geschke, Y.~Cai, J.~West,
  D.~Newth, T.~Baynes, M.~Lenzen, A.~Owen, Decoupling global environmental
  pressure and economic growth: scenarios for energy use, materials use and
  carbon emissions, Journal of cleaner production 132 (2016) 45--56.

\bibitem{fichert2020aviation}
F.~Fichert, P.~Forsyth, H.-M. Niemeier, Aviation and Climate Change: Economic
  Perspectives on Greenhouse Gas Reduction Policies, Routledge, 2020.

\bibitem{icao_proj}
I.~C.~A. Organization,
  \href{https://www.icao.int/Meetings/FutureOfAviation/Pages/default.aspx}{World
  data on aviation},
  https://www.icao.int/Meetings/FutureOfAviation/Pages/default.aspx (2020).
\newline\urlprefix\url{https://www.icao.int/Meetings/FutureOfAviation/Pages/default.aspx}

\bibitem{atag_waypoint}
A.~T.~A. Group,
  \href{https://aviationbenefits.org/media/167187/w2050_full.pdf}{Waypoint
  2050}, https://aviationbenefits.org/media/167187/w2050_full.pdf (2020).
\newline\urlprefix\url{https://aviationbenefits.org/media/167187/w2050_full.pdf}

\bibitem{lee2010can}
J.~J. Lee, Can we accelerate the improvement of energy efficiency in aircraft
  systems?, Energy conversion and management 51~(1) (2010) 189--196.

\bibitem{sanchez2017generation}
F.~Sanchez, G{\'e}n{\'e}ration de mod{\`e}les analytiques pour la conception
  pr{\'e}liminaire de syst{\`e}mes multi-physiques: application {\`a} la
  thermique des actionneurs et des syst{\`e}mes {\'e}lectriques embarqu{\'e}s,
  Ph.D. thesis, Université de Toulouse (2017).

\bibitem{kucharavy2011application}
D.~Kucharavy, R.~De~Guio, Application of s-shaped curves, Procedia Engineering
  9 (2011) 559--572.

\bibitem{jarne2007s}
G.~Jarne, J.~Sanchez-Choliz, F.~Fatas-Villafranca, “s-shaped” curves in
  economic growth. a theoretical contribution and an application, Evolutionary
  and Institutional Economics Review 3~(2) (2007) 239--259.

\bibitem{iata}
I.~A.~T. Association, \href{https://www.iata.org/}{World data on aviation},
  https://www.iata.org/ (2020).
\newline\urlprefix\url{https://www.iata.org/}

\bibitem{khandelwal2013hydrogen}
B.~Khandelwal, A.~Karakurt, P.~R. Sekaran, V.~Sethi, R.~Singh, Hydrogen powered
  aircraft: The future of air transport, Progress in Aerospace Sciences 60
  (2013) 45--59.

\bibitem{noppel2007overview}
F.~Noppel, R.~Singh, Overview on contrail and cirrus cloud avoidance
  technology, Journal of Aircraft 44~(5) (2007) 1721--1726.

\bibitem{gierens2008review}
K.~M. Gierens, L.~Lim, K.~Eleftheratos, A review of various strategies for
  contrail avoidance, Open Atmospheric Science Journal 2 (2008) 1--7.

\bibitem{teoh2020mitigating}
R.~Teoh, U.~Schumann, A.~Majumdar, M.~E. Stettler, Mitigating the climate
  forcing of aircraft contrails by small-scale diversions and technology
  adoption, Environmental Science \& Technology 54~(5) (2020) 2941--2950.

\bibitem{matthes2020climate}
S.~Matthes, B.~L{\"u}hrs, K.~Dahlmann, V.~Grewe, F.~Linke, F.~Yin,
  E.~Klingaman, K.~P. Shine, Climate-optimized trajectories and robust
  mitigation potential: Flying atm4e, Aerospace 7~(11) (2020) 156.

\bibitem{marquart2005upgraded}
S.~Marquart, M.~Ponater, L.~Str{\"o}m, K.~Gierens, An upgraded estimate of the
  radiative forcing of cryoplane contrails, Meteorologische Zeitschrift 14~(4)
  (2005) 573--582.

\bibitem{iata_covid}
I.~A.~T. Association,
  \href{https://www.iata.org/contentassets/6dfc19c3fdce4c9c8d5f1565c472b53f/2020-09-29-02-fr.pdf}{Communiqué
  de presse - prévision de trafic réduite après un été morose},
  https://www.iata.org/contentassets/6dfc19c3fdce4c9c8d5f1565c472b53f/2020-09-29-02-fr.pdf
  (2020).
\newline\urlprefix\url{https://www.iata.org/contentassets/6dfc19c3fdce4c9c8d5f1565c472b53f/2020-09-29-02-fr.pdf}

\bibitem{matthews2009proportionality}
H.~D. Matthews, N.~P. Gillett, P.~A. Stott, K.~Zickfeld, The proportionality of
  global warming to cumulative carbon emissions, Nature 459~(7248) (2009)
  829--832.

\bibitem{friedlingstein2014persistent}
P.~Friedlingstein, R.~M. Andrew, J.~Rogelj, G.~P. Peters, J.~G. Canadell,
  R.~Knutti, G.~Luderer, M.~R. Raupach, M.~Schaeffer, D.~P. van Vuuren, et~al.,
  Persistent growth of co 2 emissions and implications for reaching climate
  targets, Nature geoscience 7~(10) (2014) 709--715.

\bibitem{rogelj2016differences}
J.~Rogelj, M.~Schaeffer, P.~Friedlingstein, N.~P. Gillett, D.~P. Van~Vuuren,
  K.~Riahi, M.~Allen, R.~Knutti, Differences between carbon budget estimates
  unravelled, Nature Climate Change 6~(3) (2016) 245--252.

\bibitem{rogelj2019estimating}
J.~Rogelj, P.~M. Forster, E.~Kriegler, C.~J. Smith, R.~S{\'e}f{\'e}rian,
  Estimating and tracking the remaining carbon budget for stringent climate
  targets, Nature 571~(7765) (2019) 335--342.

\bibitem{onu_ghg}
ONU, \href{https://www.unenvironment.org/emissions-gap-report-2020}{Emissions
  gap report 2020}, https://www.unenvironment.org/emissions-gap-report-2020
  (2020).
\newline\urlprefix\url{https://www.unenvironment.org/emissions-gap-report-2020}

\end{thebibliography}







\end{document}